\documentclass[10pt]{article}
\usepackage[latin1]{inputenc}
\usepackage{amsmath}
\usepackage{amsfonts}
\usepackage{amssymb}
\usepackage{mathrsfs}
\usepackage{lineno}

\usepackage{MnSymbol}

\usepackage[cal=boondox,scr=boondoxo]{mathalfa}

\usepackage{float}
\usepackage{subfig}
\usepackage{fancybox,graphicx}
\usepackage{subfig}
\usepackage{caption}
\usepackage{color}
\usepackage{authblk}

\usepackage[colorlinks]{hyperref}

\usepackage{hyperref}

\newcommand{\doi}[1]{{doi:~\href{https://doi.org/#1}{\nolinkurl{#1}}}\rmFullStop}

\newcommand*{\rmFullStop}{\rmifnextchar{.}{}{}}

\makeatletter
\newcommand{\rmifnextchar}[3]{%
  \begingroup
  \ltx@LocToksA{\endgroup#2}%
  \ltx@LocToksB{\endgroup#3}%
  \ltx@ifnextchar{#1}{%
    \def\next{\the\ltx@LocToksA}%
    \afterassignment\next
    \let\scratch= %
  }{%
    \the\ltx@LocToksB
  }%
}
\makeatother 

\usepackage{accents}
\usepackage[titletoc,title]{appendix}
\usepackage{cite}

\usepackage{mathtools}

\usepackage[top=2in, bottom=1.5in, left=1in, right=1in]{geometry}

\definecolor{myblue}{rgb}{0.0, 0.0, 0.8}
\newcommand{\newtext}[1]{{\color{black}#1\color{black}}}

\usepackage{tikz,xcolor}

\definecolor{lime}{HTML}{A6CE39}
\DeclareRobustCommand{\orcidicon}{%
	\begin{tikzpicture}
	\draw[lime, fill=lime] (0,0) 
	circle [radius=0.16] 
	node[white] {{\fontfamily{qag}\selectfont \scriptsize ID}};
	\draw[white, fill=white] (-0.0625,0.095) 
	circle [radius=0.007];
	\end{tikzpicture}
	\hspace{-2mm}
}

\foreach \x in {A, ..., Z}{%
	\expandafter\xdef\csname orcid\x\endcsname{\noexpand\href{https://orcid.org/\csname orcidauthor\x\endcsname}{\noexpand\orcidicon}}
}

\title{Modeling of Electromagnetic Radiation using a Dual Four-Potential Representation: From Dipole Blade Radiators to Ribbon Loop-like Antennas}

\author[1,2]{Robert Salazar\orcidA{}\footnote{Corresponding author} \thanks{\href{mailto:rp.salazar84@uniandes.edu.co}{rp.salazar84@uniandes.edu.co}}}
\author[3]{Camilo Bayona-Roa\orcidB{} \thanks{\href{mailto:cabayonar@unal.edu.co}{cabayonar@unal.edu.co}}}
\affil[1]{Direcci\'on de Electr\'onica, Universidad ECCI, Bogot\'a, Colombia}
\affil[2]{Universidad Distrital Francisco Jos\'e de Caldas, Bogot\'a, Colombia}
\affil[3]{Centro de Ingenier\'ia Avanzada Investigaci\'on y Desarrollo --- CIAID, Bogot\'a, Colombia}

\begin{document}

    \maketitle

\begin{abstract}
In this {\color{black}paper, we explore classical electromagnetic radiation using} a dual four-dimensional potential $\Theta^\mu$ {\color{black}approach. Our} focus {\color{black}is} on the Planar Dipole Blade Antenna (PDBA), a system {\color{black}consisting of two flat conductive} regions on the $xy$-plane{\color{black}, separated by a gap $\mathcal{G}$, with} alternating potentials {\color{black} applied to the conductors. This method emphasizes} the use of the scalar magnetic potential $\Psi(\boldsymbol{r},t)$ and the electric vector potential $\boldsymbol{\Theta}${\color{black}, which generates} the electric field $\boldsymbol{E}(\boldsymbol{r},t)=\nabla\times\boldsymbol{\Theta}(\boldsymbol{r},t)$ in free space{\color{black}. These potentials replace} the standard magnetic vector potential $\boldsymbol{A}$ and the scalar electric potential $\boldsymbol{\Phi}$ {\color{black}in our analysis.} For harmonic radiation, the electromagnetic field can be {\color{black}expressed} in terms of the electric vector potential $\boldsymbol{\Theta}(\boldsymbol{r},t)${\color{black}. We derive} a corresponding retarded vector potential for $\boldsymbol{\Theta}$ in terms of a two-dimensional vector field $\boldsymbol{\mathcal{W}}(\boldsymbol{r},t)$, which {\color{black}flows through} the gap region $\mathcal{G}$. {\color{black}This} dual analytical approach {\color{black}yields} mathematically equivalent expressions for modeling Planar Blade Antennas, analogous to those used for ribbons {\color{black}in} the region $\mathcal{G}$, simplifying the mathematical problem. In the gapless limit, this approach {\color{black}reduces the} two-dimensional radiator (PDBA) to a one-dimensional wire-loop-like antenna{\color{black}, significantly simplifying} the {\color{black}problem's dimensionality}. This {\color{black}leads to} a dual version of Jefimenko's equations for the electric field, where $\boldsymbol{\mathcal{W}}$ {\color{black}behaves like a} surface current in the gap region {\color{black}and satisfies} a continuity condition. {\color{black}To demonstrate the utility of this approach, we provide an analytical solution for} a PDBA with a thin annular gap at low frequency.\footnote{\textit{This is the version of the article before peer review or editing, as submitted by an author to Physica Scripta. IOP Publishing Ltd is not responsible for any errors or omissions in this version of the manuscript or any version derived from it. The Version of Record is available online at} \url{https://doi.org/10.1088/1402-4896/ad98ce}}

\textbf{Keywords:} Blade antennas, Dual Jefimenko's Equations, Electric Vector Potential, Gapless and Gapped Time-Dependent Surface Electrodes.
\end{abstract}

\section{Introduction}
The system of interest in {\color{black}this study consists of} two conductive sheets: $\mathcal{A}_{in}$ and $\mathcal{A}_{out}$, {\color{black}both} lying on the $xy$-plane and separated by a gap, {\color{black}defined} as follows:
\begin{equation}
\mathcal{G} = \left\{ (r,\phi,0) : \hspace{0.25cm} \mathscr{R}(\phi) - \frac{\nu}{2} < r < \mathscr{R}(\phi) + \frac{\nu}{2}, \hspace{0.25cm} \forall \hspace{0.25cm} \phi \in [0,2\pi) \right\},    
\label{gapEq}
\end{equation}
where $\mathscr{R}(\phi)$, $(\mathscr{R}(\phi) - \nu/2, \phi)$, and $(\mathscr{R}(\phi) + \nu/2, \phi)$ {\color{black}represent} the parametric {\color{black}contours} of $\partial A$, $\partial \mathcal{A}_{in}$, and $\partial \mathcal{A}_{out}$, respectively. These surfaces satisfy the condition $\mathcal{A}_{in} \cup \mathcal{G} \cup \mathcal{A}_{out} = \mathbb{R}^2$. 

\begin{figure}[H]
\centering
\includegraphics[width=0.65\textwidth]{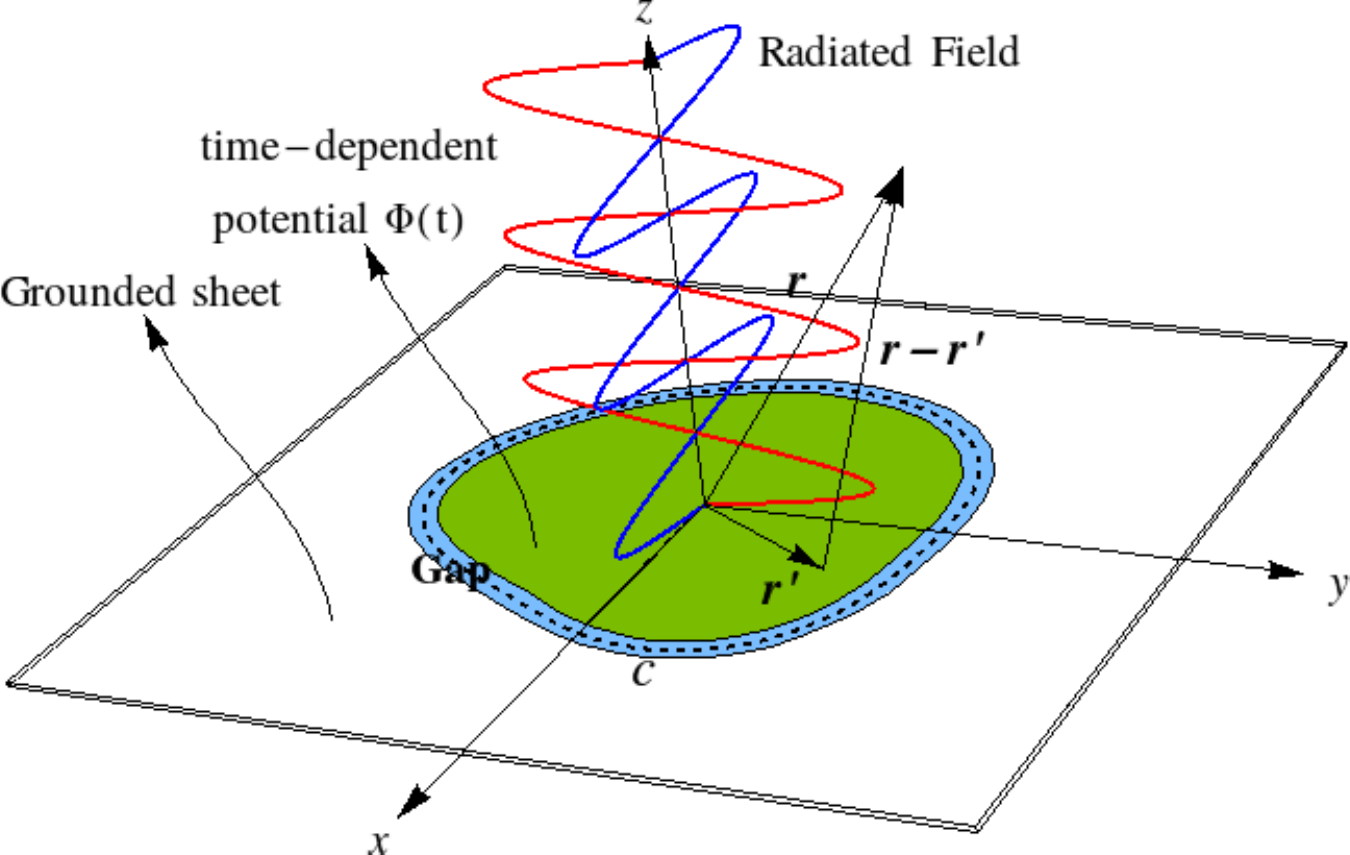}
\caption[System schematic]{Schematic of the system: Two conductive sheets, $\mathcal{A}_{\text{in}}$ and $\mathcal{A}_{\text{out}}$, separated by a gap $\mathcal{G}$. The inner sheet $\mathcal{A}_{\text{in}}$ is subjected to a time-dependent potential, $V(t)$, while the outer sheet $\mathcal{A}_{\text{out}}$ is grounded.}
\label{theSystemFig}
\end{figure}

For {\color{black}clarity,} the two-dimensional plane {\color{black}is expressed in} Cartesian {\color{black}coordinates} as $\mathbb{R}^2 = \{(x,y,0)\}$. The finite inner region $\mathcal{A}_{in}$ {\color{black}is characterized by a time-varying} scalar potential $\Phi_o(t) = V(t)$, while the outer metallic layer $\mathcal{A}_{out}$ is grounded (see Fig.~\ref{theSystemFig}). 

Under these conditions, the system {\color{black}radiates} electromagnetic waves {\color{black}into} the $\mathbb{R}^3$ space. {\color{black}Our focus is on the radiation} in the half-space region $\mathfrak{D} = \left\{ \boldsymbol{r} \in \mathbb{R}^3 : z > 0 \right\}$, as the system exhibits mirror symmetry with respect to the $z=0$ {\color{black}plane. This mirror symmetry arises because the electric charges are confined to the conductive layers on} the $xy$-plane.

The system of interest is {\color{black}configured as} a Planar Dipole Blade Antenna (PDBA) radiating in free space. {\color{black}This technology is crucial for} various technical applications, including communications, navigation, aeronautics \cite{macnamara2010introduction}, and astronomy \cite{hicks2012wide,mozdzen2016limits}. Theoretical modeling of antenna devices {\color{black}has garnered significant interest, particularly in} wire-antenna radiation setups, where wires are driven by alternating voltage sources. Other key studies focus on circular loop antennas \cite{foster1944loop, storer1956impedance, lindsay1960circular, wu1962theory, werner1996exact, li1999method, li2004method}, {\color{black}utilizing} techniques such as the method of moments \cite{anastassiu2006fast,gibson2021method} and {\color{black}basis function expansions} \cite{hamed2013exact}. In general, wire antennas are mathematically simpler than PDBAs due to the spatial localization of currents.  \newtext{Other key studies focus on the design and fabrication of antennas, such as the Pattern Director Antenna (PDA), which simplifies mode conversion while achieving high directivity  \cite{Hosseini_Oraizi_2022} and the Smile-like Mode Converter Antenna (SMCA), designed for high-power applications to enhance the efficiency and power capacity of the device \cite{Hosseini2021}}. 

\newtext{Dielectric antennas have attracted attention for their high radiation efficiency, compact size, and lightweight structure. Unlike dipole blade antennas, they use dielectric materials to guide and radiate electromagnetic waves. An example is the dielectric resonator antenna (DRA), which enhances radiation by exciting resonance modes within the material. Recent innovations, like printable planar dielectric antennas, achieve high efficiency and broadside patterns with low cross-polarization at Ku-band frequencies \cite{7332737}. DRAs have advanced in multi-resonance, broadband, and ultra-wideband designs, improving bandwidth and stability \cite{soren2014dielectric}. Despite structural differences from dipole-blade antennas, dielectric antennas offer valuable advantages across modern applications \cite{623123}. Nano-antennas, operating at optical frequencies, extend this concept in nanotechnology to interact with light at subwavelength scales. Unlike traditional antennas, they rely on plasmonic \cite{alu2008tuning} or dielectric materials to enable strong light-matter interactions, making them ideal for applications in sensing and optical communications. Additionally, dielectric nano-antennas provide enhanced field confinement and avoid metallic losses \cite{Arseniy2016}. These capabilities at the intersection of photonics and electronics contrast to both dipole-blade and dielectric antennas\cite{sato2012strong}.}

The main goal {\color{black}of this} paper is to model PDBA radiation by formulating the dynamic phenomena {\color{black}through} the vector potential $\boldsymbol{\Theta}(\bold{r},t)$, {\color{black}where} the electric field is {\color{black}expressed as} $\boldsymbol{E}=\nabla\times\boldsymbol{\Theta}$. {\color{black}Using this approach}, we demonstrate that PDBAs, {\color{black}such as} the one depicted in Fig.~\ref{theSystemFig}, can be {\color{black}effectively mapped to} a wire loop-like antenna counterpart (see Fig.~\ref{ribbonFig}) in the half-space $\mathcal{D}$. This is a generalization of the formalism {\color{black}developed} for Surface Electrodes (SE) \cite{salazar2020gaped, salazar2019AngularDependentSE, romero2022monte}, which represents the electrostatic version of the system {\color{black}shown} in Fig.~\ref{theSystemFig}. In that {\color{black}context}, the electric field {\color{black}is} represented by a Biot-Savart-like integral, associating a vector potential with the electric field {\color{black}instead of} a scalar potential.

The {\color{black}concept of the} electric vector potential is not new. {\color{black} In magnetostatics, a}ssociating a scalar magnetic potential to the magnetic field is standard {\color{black}practice. However, associating a} vector potential to the electric field is {\color{black}less} common but {\color{black}applicable in various} scenarios \cite{stark2015boundary, semenov2006vector, landis2002new, albanese1990treatment}. A {\color{black}classic} example is the study of electromagnetic radiation in the Coaxial Line onto a Ground Plane (CLGP) {\color{black}system}. There, the electric field $\boldsymbol{E}$ in the transition line is {\color{black}analogous to that} of an infinitely long cylindrical capacitor. The radiated field of the CLGP can be approximated by a \textit{magnetic current field} circulating {\color{black}around} the aperture (the coaxial opening), {\color{black}following} the same profile {\color{black}as} $\boldsymbol{E}$ \cite{fields1961time, levine1951theory}. The {\color{black}approach of identifying} two different systems {\color{black}that share} the same mathematical {\color{black}structure, allowing their} solutions {\color{black}to} be interchanged, is commonly referred to as the duality theorem (see \cite{balanis2016antenna}, Chapter 3, Section 3.7).

The goal of {\color{black}this work} is to rigorously establish the mathematical equivalence between the PDBA and {\color{black}a ribbon-like} antenna {\color{black}subjected} to a specific surface {\color{black}current} density. {\color{black}This is particularly important} since the fields in the gap region are generally unknown \textit{a priori}. {\color{black}For} the magnetostatic counterpart in the low-frequency limit, {\color{black}the electric current has an} electrostatic counterpart: the Surface Electrode (SE), assuming the same gap and ribbon geometries. To achieve this, {\color{black}we must properly} define the Green's function for the problem{\color{black}, ensuring the} correct {\color{black}inclusion of the system's} boundary conditions. {\color{black}The contribution of the paper is the formal demonstration of the mathematical equivalency in the specific antenna setting. The demonstration is required to avoid mismatching and in general, it is laborious since it is necessary to build properly the Green's function of the dual potentials to fit correctly the boundary conditions, as it has been done for the PDBA. The paper also exposes a numeric-analytic strategy to approximate $\mathcal{W}(\boldsymbol{r},t)$ in the gap region.}

This document is organized as follows. 
In Section 2, {\color{black}we describe} the electric vector potential $\boldsymbol{\Theta}(\boldsymbol{r},t)$ and the dual 4-electromagnetic potential. {\color{black}By} the end of that section, we {\color{black}derive} the wave equation in free space for {\color{black}both} the electric vector potential and the magnetic scalar potential. 
Section 3 is {\color{black}dedicated} to finding an integral expression for the electric vector potential by solving the wave equation. In {\color{black}this} section, we {\color{black}also discuss the appropriate} boundary conditions on the metallic sheets and the gap.
{\color{black}In Section 4, the} radiated electric field is derived{\color{black}, and we establish a} correspondence with Jefimenko's equations. {\color{black}We} demonstrate that PDBA radiation is mathematically equivalent to the radiation {\color{black}from} a ribbon loop-like {\color{black}structure} of alternating current {\color{black}located in} the PDBA gap region.
Section 5 {\color{black}provides} expressions for the electromagnetic field in the {\color{black}specific} case of harmonic radiation.
An example {\color{black}applying} the analytic strategy to {\color{black}th}e low-frequency radiation of a circular planar dipole blade antenna is presented in Section 6. In Section 6.1, we test {\color{black}the consistency of the system with} dipole radiation in the{\color{black} far-field limit. Additionally, we} present computations of the fields near the gap as a function of the gap thickness at {\color{black}low frequencies.
Finally, the conclusions are summarized} in Section 7.

\section{The electromagnetic dual 4-potential vector representation}

\newtext{In the current section} we {\color{black}begin by outlining the strong form of the problem, which provides a direct description of the field equations. This is followed by a brief review of the standard electromagnetic 4-potential vector, which allows for the transformation of the strong form into an integral form involving the field sources: charges and currents.

The core of this work focuses on} the application of a dual electromagnetic 4-potential vector representation. {\color{black}Thus,} the treatment of the wave propagation problem is {\color{black}discussed in detail} in the remaining part of this section.




\subsection{Standard electromagnetic 4-potential vector}

A standard {\color{black}approach} to derive the electric and magnetic fields is {\color{black}through} the electromagnetic 4-potential:
\begin{equation}
\boldsymbol{A}^\mu(\boldsymbol{r},t) := \left(\frac{1}{c}\Phi(\boldsymbol{r},t), \boldsymbol{A}(\boldsymbol{r},t) \right),  
\label{fourPotentialEq}
\end{equation}
which satisfies the {\color{black}following} vector identities:
\[
\boldsymbol{E}(\boldsymbol{r},t) = -\nabla\Phi(\boldsymbol{r},t) - \frac{\partial \boldsymbol{A}(\boldsymbol{r},t)}{\partial t} \hspace{0.5cm}\text{and}\hspace{0.5cm} \boldsymbol{B}(\boldsymbol{r},t) = \nabla \times \boldsymbol{A}(\boldsymbol{r},t),
\]
where $\Phi$ is the scalar electric potential and $\boldsymbol{A}$ is the magnetic vector potential. 

In general, these fields have integral solutions that account for their spatial sources:
\begin{equation}
\Phi(\boldsymbol{r},t) = \frac{1}{4\pi\epsilon_0} \int_{\mathbb{R}^3} \frac{\rho(\boldsymbol{r}',t_{\boldsymbol{r}})}{|\boldsymbol{r} - \boldsymbol{r}'|} \, d^3\boldsymbol{r}', \label{phiIntegralEq}
\end{equation}

\begin{equation}
\boldsymbol{A}(\boldsymbol{r},t) = \frac{\mu_0}{4\pi} \int_{\mathbb{R}^3} \frac{\boldsymbol{J}(\boldsymbol{r}',t_{\boldsymbol{r}})}{|\boldsymbol{r} - \boldsymbol{r}'|} \, d^3\boldsymbol{r}', \label{AIntegralEq}
\end{equation}
with $t_{\boldsymbol{r}}$ being the retarded time.

These integral formulas are useful for wire-radiation problems, including low-frequency radiation, where {\color{black}the charge density and current density can be assumed to be} uniformly distributed. In the low-frequency case, $\boldsymbol{J}$ is {\color{black}typically well-localized along} the infinitely thin wire path. However, the {\color{black}application of Eqs.~(\ref{phiIntegralEq}) and (\ref{AIntegralEq}) becomes more} challenging in surface radiation scenarios. This {\color{black}difficulty arises because} sources, both \textit{charges and currents}, are{\color{black} generally not uniformly distributed over} the antenna surfaces. 

Even in the electrostatic versions of the surface system \cite{salazar2020gaped, romero2022monte}, the {\color{black}distribution of} surface charge density on the metallic sheets strongly {\color{black}depends} on the gap geometry. Therefore, the stationary {\color{black}form} of Eq.~(\ref{phiIntegralEq}) is not {\color{black}applicable, as} the charge density sources are \textit{a priori} unknown.

\subsection{Dual electromagnetic 4-potential vector}
The mathematical electrostatic problem can be simplified by associating a \textit{vector potential} $\boldsymbol{\Theta}$ {\color{black}with} the electric field instead of the scalar potential $\Phi$. Our goal here is to generalize the electrostatic strategy introduced in \cite{salazar2020gaped, romero2022monte} to time-dependent radiation problems.

In free-charge space, where {\color{black}both the current} density $\boldsymbol{J}(\boldsymbol{r},t) = 0$ {\color{black}and the charge density} $\rho(\boldsymbol{r},t) = 0$, the electric field $\boldsymbol{E}(\boldsymbol{r},t)$ becomes divergenceless{\color{black}, i.e.,} $\nabla \cdot \boldsymbol{E}(\boldsymbol{r},t) = 0$, allowing us to express it as:
\begin{equation}
\boldsymbol{E}(\boldsymbol{r},t) = \nabla \times \boldsymbol{\Theta}(\boldsymbol{r},t), 
\label{electricFieldIdentityEq}
\end{equation}
where $\boldsymbol{\Theta}(\boldsymbol{r},t)$ is any vector potential satisfying {\color{black}this} identity. This \textit{electric vector potential} plays a role {\color{black}analogous to the standard magnetic} vector potential $\boldsymbol{A}(\boldsymbol{r},t)${\color{black}, which is related to} the magnetic field {\color{black}via} the curl operation $\boldsymbol{B}(\boldsymbol{r},t) = \nabla \times \boldsymbol{A}(\boldsymbol{r},t)$. We can combine Eq.~(\ref{electricFieldIdentityEq}) with Ampere-Maxwell's law
\[
\nabla \times \boldsymbol{B}(\boldsymbol{r},t) = \mu_0 \boldsymbol{J}(\boldsymbol{r},t) + \mu_0 \epsilon_0 \frac{\partial \boldsymbol{E}(\boldsymbol{r},t)}{\partial t}
\]
in vacuum ($\boldsymbol{J}=0$), {\color{black}leading to:}
\[
\nabla \times \boldsymbol{B}(\boldsymbol{r},t) = \frac{1}{c^2} \frac{\partial }{\partial t}\left(\nabla \times \boldsymbol{\Theta}(\boldsymbol{r},t)\right) \hspace{0.5cm} \therefore \hspace{0.5cm} \nabla \times \left( \boldsymbol{B}(\boldsymbol{r},t) - \frac{\partial }{\partial t} \frac{\boldsymbol{\Theta}(\boldsymbol{r},t)}{c^2} \right) = 0.
\]
Since the curl of the gradient of any scalar field is zero, we can express the term {\color{black}inside the parentheses as the} gradient of an arbitrary scalar field, say $\nabla(-\mu_0\Psi(\boldsymbol{r},t))$. {\color{black}Here,} $\Psi(\boldsymbol{r},t)$ {\color{black}is identified} as the \textit{magnetic scalar potential}, {\color{black}a convenient choice that} maintains dimensional consistency. Thus, the magnetic field takes the form:
\begin{equation}
\boldsymbol{B}(\boldsymbol{r},t) = -\mu_0 \nabla \Psi(\boldsymbol{r},t) + \frac{\partial }{\partial t} \frac{\boldsymbol{\Theta}(\boldsymbol{r},t)}{c^2}.    
\label{magneticFieldIdentityEq}
\end{equation}

{\color{black}Given} Eqs.~(\ref{electricFieldIdentityEq}) and (\ref{magneticFieldIdentityEq}), it is {\color{black}useful} to compose a dual 4-potential vector by {\color{black}organizing the terms as:}
\[
\boldsymbol{\Theta}^\mu(\boldsymbol{r},t) := \left( a_0 \mu_0 \Psi(\boldsymbol{r}), a_1 \frac{\Theta_x(\boldsymbol{r},t)}{c^2}, a_1 \frac{\Theta_y(\boldsymbol{r},t)}{c^2}, a_1 \frac{\Theta_z(\boldsymbol{r},t)}{c^2} \right),
\]
where the first term {\color{black}corresponds} to the magnetic scalar potential and the {\color{black}subsequent terms include the components of the} electric vector potential. Here, $a_0$ and $a_1$ are constants {\color{black}specific to} the problem. We adopt physical coordinates as $\left\{ x^\mu \right\}_{\mu=0,\ldots,3} = \left\{ ct, x, y, z \right\}$.

Since {\color{black}$\boldsymbol{\Theta}'(\boldsymbol{r},t) = \boldsymbol{\Theta}(\boldsymbol{r},t) + \nabla\lambda(\boldsymbol{r},t)$}, {\color{black}where} $\lambda$ {\color{black}is} an arbitrary scalar function, the vector potential in Eq.~(\ref{electricFieldIdentityEq}) is not unique. Therefore, a {\color{black}divergence-free} condition must be imposed {\color{black}on} the 4-potential vector, {\color{black}taking} the form:
\[
\partial_\mu \boldsymbol{\Theta}^\mu(\boldsymbol{r},t) = \boldsymbol{\Theta}^\mu_{,\mu} (\boldsymbol{r},t) = 0,
\]
which resembles a Lorentz {\color{black}gauge-like} condition. Here, we adopt the Einstein summation convention for repeated indices, with $\partial_\mu$ denoting a covariant derivative. \footnote{This is essential {\color{black}for selecting an appropriate gauge} condition. {\color{black}For time-steady problems, this condition reduces to} $\nabla \cdot \boldsymbol{\Theta} = 0${\color{black}, similar to the} Coulomb gauge. This gauge is useful {\color{black}for modeling} electrostatic systems via the electric vector potential, as shown in \cite{salazar2020gaped, romero2022monte} for surface electrodes and Coulomb two-dimensional gases.}

Finally, we demonstrate that the dual 4-potential vector is invariant under space-time transformations{\color{black}, as detailed} in Appendix \ref{Lorentzappendix}.

\subsubsection{D'Alembert equation for the magnetic scalar potential}

Rearranging the summation result, we find the following relation between the temporal term and the spatial divergence:
\begin{equation}
\nabla \cdot \left( \frac{1}{c^2} \boldsymbol{\Theta}(\boldsymbol{r},t) \right) =  -\frac{1}{a_1} \frac{\partial}{\partial (ct)} \left( a_0 \mu_0 \Psi(\boldsymbol{r},t) \right). 
\label{aux1Eq}
\end{equation}

Additionally, we can {\color{black}apply} Gauss's law for the magnetic field to Eq.~(\ref{magneticFieldIdentityEq}):
\[
\nabla \cdot \boldsymbol{B}(\boldsymbol{r},t) = -\mu_0 \nabla^2 \Psi(\boldsymbol{r},t) + \frac{\partial}{\partial t} \frac{\nabla \cdot \boldsymbol{\Theta}(\boldsymbol{r},t)}{c^2} = 0,
\]
to obtain the constants $a_0$ and $a_1$. {\color{black}By combining this with Eq.~(\ref{aux1Eq}), we have:}
\[
\nabla^2 \left( \mu_0 \Psi(\boldsymbol{r},t) \right) + \frac{\partial}{\partial t} \left[ \frac{1}{a_1} \frac{\partial}{\partial (ct)} \left( a_0 \mu_0 \Psi(\boldsymbol{r},t) \right) \right] = 0.
\]

{\color{black}Assuming $a_0 = -a_1 / c$,} this expression simplifies to:
\begin{equation}
    \Box^2 \Psi(\boldsymbol{r},t) = 0, \hspace{0.5cm} \text{where} \hspace{0.5cm} \Box^2 = \nabla^2 - \frac{1}{c^2} \frac{\partial^2}{\partial t^2},
    \label{waveEquationPsiEq}
\end{equation}
{\color{black}with} $\Box^2$ {\color{black}representing the D'Alembertian operator.}

We are still free to choose the value of one of the constants. If we set $a_1 = -1$, then $a_0 = 1/c$, and the dual version of the electromagnetic potential takes the form:
\begin{equation}
\boldsymbol{\Theta}^\mu(\boldsymbol{r},t) := \left( \frac{\mu_0}{c} \Psi(\boldsymbol{r},t), -\frac{\boldsymbol{\Theta}(\boldsymbol{r},t)}{c^2} \right),  
\label{dual4PotentialEq}
\end{equation}
under the following Lorentz gauge-like condition:
\begin{equation}
\nabla \cdot \boldsymbol{\Theta}(\boldsymbol{r},t) - \mu_0 \frac{\partial \Psi(\boldsymbol{r},t)}{\partial t} = 0.    
\label{lorenztGaugeLikeConditionEq}
\end{equation}

\subsubsection{D'Alambert equation for the dual 4-potential vector}
Now, we can {\color{black}substitute} Eq.~(\ref{electricFieldIdentityEq}) {\color{black}into} Faraday's law:
\[
\nabla \times \boldsymbol{E}(\boldsymbol{r},t) = \nabla \times \nabla \times \boldsymbol{\Theta}(\boldsymbol{r},t) = - \frac{\partial \boldsymbol{B}(\boldsymbol{r},t)}{\partial t}.
\]
By applying the vector calculus identity $\nabla^2 \boldsymbol{F}(\boldsymbol{r}) = \nabla (\nabla \cdot \boldsymbol{F}(\boldsymbol{r})) - \nabla \times \nabla \times \boldsymbol{F}(\boldsymbol{r})$ for any vector field $\boldsymbol{F}(\boldsymbol{r})$, we obtain:
\[
-\nabla^2 \boldsymbol{\Theta}(\boldsymbol{r},t) + \nabla (\nabla \cdot \boldsymbol{\Theta}(\boldsymbol{r},t)) = - \frac{\partial \boldsymbol{B}(\boldsymbol{r},t)}{\partial t}.
\]
{\color{black}Next, using} the Lorentz-like gauge condition from Eq.~(\ref{aux1Eq}) with $(a_0,a_1) = (1/c, -1)${\color{black}, we simplify the equation} to:
\[
-\nabla^2 \boldsymbol{\Theta}(\boldsymbol{r},t) + \nabla \left(-\mu_0 \frac{\partial}{\partial t} \Psi(\boldsymbol{r},t) \right) = - \frac{\partial \boldsymbol{B}(\boldsymbol{r},t)}{\partial t}.
\]
This result can {\color{black}be further rearranged} as:
\[
-\nabla^2 \boldsymbol{\Theta}(\boldsymbol{r},t) = -\frac{\partial}{\partial t} \left( \boldsymbol{B}(\boldsymbol{r},t) + \mu_0 \nabla \Psi(\boldsymbol{r},t) \right) = -\frac{\partial}{\partial t} \left( \frac{\boldsymbol{\Theta}(\boldsymbol{r},t)}{c^2} \right),
\]
where we have used the relation from Eq.~(\ref{magneticFieldIdentityEq}). {\color{black} Thus,} we obtain the wave equation for the dual 4-potential vector:
\begin{equation}
\left( -\nabla^2 + \frac{1}{c^2} \frac{\partial^2}{\partial t^2} \right) \boldsymbol{\Theta}(\boldsymbol{r},t) = \Box^2 \boldsymbol{\Theta}(\boldsymbol{r},t) = 0.
\label{waveEquationThetaEq}
\end{equation}

\subsubsection{Generalized Poisson's equations}

Equations~(\ref{waveEquationPsiEq}) and (\ref{waveEquationThetaEq}) can be combined and written in {\color{black}a more compact} form:
\begin{equation}
    \Box^2 \boldsymbol{\Theta}^\mu(\boldsymbol{r},t) = 0,
    \label{Laplace4-potential}
\end{equation}
with the introduction of 
$
\Box^2 = \eta^{\mu \nu} \partial_\mu \partial_\nu,
$
{\color{black}where} we adopt $\eta := \bold{diag}(1, -1, -1, -1)$ as the Minkowski metric{\color{black}. This} implies that the dual 4-potential vector $\boldsymbol{\Theta}^\mu(\boldsymbol{r},t)$ satisfies a wave equation in free-charge space, similar to its dual counterpart $\boldsymbol{A}^\mu(\boldsymbol{r},t)$ {\color{black}as} given by Eq.~(\ref{fourPotentialEq}).

This dual representation resembles a Laplace-type equation. {\color{black}Furthermore,} the Laplace problem {\color{black}can be generalized} as a Poisson problem with zero source terms. Thus, we generalize {\color{black}the problem in Eq.~}(\ref{waveEquationThetaEq}) into a 4-dimensional Poisson equation of the form:
\begin{equation}
    \Box^2 \boldsymbol{\Theta}^\mu(\boldsymbol{r},t) = \boldsymbol{S}^\mu(\boldsymbol{r},t),
    \label{eq:4dimensionalPoisson}
\end{equation} 
where $\boldsymbol{S}^\mu(\boldsymbol{r},t)$ is a 4-dimensional vector representing the sources.

Appropriate boundary conditions must be set for the Dipole Blade Radiation problem described {\color{black}by} Eq.~(\ref{eq:4dimensionalPoisson}). Careful definitions of {\color{black}these} boundary conditions must be imposed{\color{black}, considering} the mixed time and spatial variables in the unknowns. 

Therefore, we {\color{black}focus on solving} the generalized Poisson problem {\color{black}for} the electric vector potential. {\color{black}This approach allows us to establish spatial boundary} conditions over the unknowns.

\section{The radiated electric vector potential}

In this section, we {\color{black}derive} the analytical expression for the electric vector potential in the $\mathbb{R}^3$ field, within the planar blade setting with the arbitrary gap $\mathcal{G}$ defined as in~(\ref{gapEq}). Several analytical solutions {\color{black}to the Poisson equation} for three-dimensional problems {\color{black}have been proposed} \cite{salazar2020gaped, salazar2022electric}, {\color{black}most of which} rely on the construction of a Green's function {\color{black}by evaluating} specific boundary conditions. {\color{black}This approach is} adopted and explained here.

\subsection{Green's function}

Analytical solutions to Eq.~(\ref{eq:4dimensionalPoisson}) can be found by introducing $G(\boldsymbol{r},t;\boldsymbol{r}',t')$ as the Green's function of the D'Alembert operator:
\begin{align*}
\boldsymbol{\Theta}(\boldsymbol{r},t) &= \int_{\mathcal{D} \times \mathbb{T}} G(\boldsymbol{r},t;\boldsymbol{r}',t') \boldsymbol{S}(\boldsymbol{r}',t') \, d^3\boldsymbol{r}' \, dt' \\
&+ \int_\mathbb{T} \oint_{\partial \mathcal{D}} \left[ \boldsymbol{\Theta}(\boldsymbol{r}',t') \frac{\partial}{\partial n'} G(\boldsymbol{r},t;\boldsymbol{r}',t') - G(\boldsymbol{r},t;\boldsymbol{r}',t') \frac{\partial}{\partial n'} \Theta(\boldsymbol{r}',t') \right] d^2\boldsymbol{r}' \, dt',
\end{align*}
with appropriate conditions for the Green's function over the domain. Here, $\mathbb{T} = \left\{ t : -\infty < t < \infty \right\}$ is the temporal domain, $\mathfrak{D} = \left\{ \boldsymbol{r} \in \mathbb{R}^3 : z > 0 \right\}$ is the spatial domain, and $\hat{n}(\boldsymbol{r}')$ represents the normal outward vector to the $\partial \mathcal{D}$ surface. {\color{black}The Green's function satisfies:}
\begin{align}
\Box^2 G_{N}(\boldsymbol{r},t;\boldsymbol{r}',t') = \delta^3(\boldsymbol{r} - \boldsymbol{r}') \delta(t - t'). \label{eq:Greenfunction}
\end{align}

Here, $\partial_n' G_{N}(\boldsymbol{r},\boldsymbol{r}') = \nabla' G_{N}(\boldsymbol{r},\boldsymbol{r}') \cdot \hat{n}(\boldsymbol{r}')$ refers to the normal derivative. In the electromagnetic radiation problem, the volumetric sources $\boldsymbol{S}(\boldsymbol{r}',t')$ are negligible, and {\color{black}only} the boundary terms contribute to the Green's function. We choose the Green's function such that {\color{black}Neumann} boundary conditions satisfy the electric vector potential as:
\[
\boldsymbol{\Theta}(\boldsymbol{r},t) = -\int_\mathbb{T} \oint_{\partial \mathcal{D}} G_{N}(\boldsymbol{r},t;\boldsymbol{r}',t') \frac{\partial}{\partial n'} \Theta(\boldsymbol{r}',t') d^2\boldsymbol{r}' \, dt'.
\]
The surface contribution vanishes on $\partial \mathcal{C} = \partial \mathcal{D} \setminus \partial \mathcal{D}_{xy}$, with $\mathcal{D}_{xy}$ being the $z = 0$ plane. This simplifies the {\color{black}expression to:}
\[
\boldsymbol{\Theta}(\boldsymbol{r},t) = \int_\mathbb{T} \int_{\mathbb{R}^2} \left. G_{N}(\boldsymbol{r},t;\boldsymbol{r}',t') \frac{\partial}{\partial z'} \Theta(\boldsymbol{r}',t') \right|_{z' = 0} dx' \, dy' \, dt'.
\]
The electric field can be calculated by approaching the sheets on the $xy$-plane from the $z$-direction:
\[
\lim_{z \to 0^+} \boldsymbol{E}(\boldsymbol{r},t) = (0,0,E_z(x,y,0,t)),
\]
which {\color{black}assumes that} tangential electric currents {\color{black}are negligible.}

\subsection{Electrode surfaces}

The electric field on the metallic sheets is computed from the curl of the vector potential:
\[
(0,0,E_z(x,y,0,t)) = \left( \partial_y \Theta_z - \partial_z \Theta_y, \partial_z \Theta_x - \partial_x \Theta_z, \partial_x \Theta_y - \partial_y \Theta_x \right)  \hspace{0.25cm} \forall \hspace{0.25cm} (x,y,0) \notin \mathcal{G}.
\]
Since the {\color{black}components} $\partial_y \Theta_z = 0$ and $\partial_x \Theta_z = 0$ must {\color{black}vanish} due to the absence of tangential currents, the electric field {\color{black}becomes:}
\[
\lim_{z \to 0^+} \boldsymbol{E}(\boldsymbol{r},t) = \left( -\partial_z \Theta_y, \partial_z \Theta_x, \partial_x \Theta_y - \partial_y \Theta_x \right)  \hspace{0.25cm} \forall \hspace{0.25cm} (x,y,0) \notin \mathcal{G}.
\]
{\color{black}Since the terms} $\partial_z \Theta_y$ and $\partial_z \Theta_x$ vanish, the dual vector potential{\color{black}, which results from} the charge distributions on the metallic sheets, is a superposition of potentials due to {\color{black}point} charges. {\color{black}Consequently,} $\Theta_z = 0$ in $\mathbb{R}^3$, and thus $\partial_z \Theta_z = 0$. {\color{black}Therefore,} in the $z \to 0$ limit, {\color{black}we have:}
\[
\lim_{z \to 0} \frac{\partial}{\partial z} \boldsymbol{\Theta}(\boldsymbol{r},t) = 0 \hspace{0.25cm} \forall \hspace{0.25cm} (x,y,0) \notin \mathcal{G}.
\]
{\color{black}This indicates} that the normal derivative of the vector potential on the $xy$-plane is zero everywhere except in the gap region.

\subsection{Gap surface}

The $z$-derivative of the dual 4-potential vector on the gap $\partial_z \boldsymbol{\Theta}(\boldsymbol{r} \in \mathcal{G},t)$ is related to the magnetic vector potential $\boldsymbol{A}$ in the following form:
\[
\partial_z \boldsymbol{\Theta}(\boldsymbol{r},t) = \left( -\partial_y \Phi + \partial_t A_y, \partial_x \Phi - \partial_t A_x, 0 \right) \hspace{0.25cm} \forall \hspace{0.25cm} (x,y,0) \in \mathcal{G}.
\]
Moreover, the vector field arising from the normal derivative of the vector potential $\partial_z \boldsymbol{\Theta}(\boldsymbol{r} \in \mathcal{G},t)$ is related to a \textit{Weight Vector} field over the gap, denoted as $\boldsymbol{\mathcal{W}}(\boldsymbol{r} \in \mathcal{G},t)$. This weight vector has been introduced and applied to the electrostatic version of the problem in \cite{salazar2020gaped, salazar2022electric}. Its mathematical definition is obtained from $\partial_z \boldsymbol{\Theta}(\boldsymbol{r} \in \mathcal{G},t) =: -\boldsymbol{\mathcal{W}}(\boldsymbol{r} \in \mathcal{G},t)$, as follows:
\[
\boldsymbol{\mathcal{W}}(\boldsymbol{r},t) := - \left( -\partial_y \Phi + \partial_t A_y, \partial_x \Phi - \partial_t A_x, 0 \right) \hspace{0.25cm} \forall \hspace{0.25cm} (x,y,0) \in \mathcal{G}.
\]

It is convenient to {\color{black}express} $\boldsymbol{\mathcal{W}}(\boldsymbol{r},t)$ in polar coordinates $\boldsymbol{r} = (u,\phi,z)$, with $(\hat{u}, \hat{\phi}, \hat{z})$ {\color{black}as the unit vectors, yielding:}
\begin{equation}
\boxed{
    \boldsymbol{\mathcal{W}}(u,\phi,t) = \left( \frac{1}{u} \frac{\partial \Phi}{\partial \phi} - \frac{\partial A_\phi}{\partial t} \right) \hat{u} + \left( -\frac{\partial \Phi}{\partial u} + \frac{\partial A_u}{\partial t} \right) \hat{\phi} \hspace{0.25cm} \forall \hspace{0.25cm} (u,\phi) \in \mathcal{G}, \hspace{0.25cm} \text{otherwise} \hspace{0.25cm} 0.
}
    \label{weightVectorEq}
\end{equation}


\subsection{Electric vector potential}

The solution to the generalized Poisson problem takes the form:
\[
\boldsymbol{\Theta}(\boldsymbol{r},t) = -\int_\mathbb{T} \int_{\mathcal{G} \in \mathbb{R}^2} \left. G_{N}(\boldsymbol{r},t;\boldsymbol{r}',t') \right|_{z'=0} \boldsymbol{\mathcal{W}}(u',\phi',t') u' \, du' \, d\phi' \, dt'.
\]
In the case of an infinite domain, the Green's function of the D'Alembert operator in Eq.~(\ref{eq:Greenfunction}) is given by:
\begin{equation}
G_N(\boldsymbol{r},t;\boldsymbol{r}',t') = -\frac{\delta(t' - t_{\boldsymbol{r}})}{4\pi |\boldsymbol{r} - \boldsymbol{r}'|}, \hspace{0.5cm} \text{where} \hspace{0.5cm} t_{\boldsymbol{r}} := t - \frac{|\boldsymbol{r} - \boldsymbol{r}'|}{c}
\label{greenFunctionNoBoundEq}
\end{equation}
is the retarded time (see Appendix \ref{greenFunctionAppendix} for a detailed explanation of this result). We impose the {\color{black}Neumann boundary condition:}
\begin{equation}
\frac{\partial}{\partial n'} G_{N}(\boldsymbol{r},t;\boldsymbol{r}',t') = 0 \hspace{0.25cm} \forall \hspace{0.25cm} \boldsymbol{r}' \in \partial \mathcal{D}
\label{NeumannCondtionOnGreenFunctionEq}
\end{equation}
to define the {\color{black}Neumann} boundary conditions on the electric vector potential $\boldsymbol{\Theta}(\boldsymbol{r},t)$. {\color{black}The} Green's function in Eq.~(\ref{greenFunctionNoBoundEq}) can be interpreted as a {\color{black}point} source with a spherical wavefront. To satisfy the boundary conditions, we use the method of images and propose $G_{N}(\boldsymbol{r}, \boldsymbol{r}')$ as the Green's function of two identical {\color{black}point} sink/source charges symmetrically placed at $z'$ and $-z'$:
\begin{equation}
    G_N(\boldsymbol{r},t;\boldsymbol{r}',t') = -\frac{1}{4\pi} \frac{\delta(t' - t_{\boldsymbol{r}})}{|\boldsymbol{r} - \boldsymbol{r}'|} + \left[ -\frac{1}{4\pi} \frac{\delta(t' - t_{\boldsymbol{r}})}{|\boldsymbol{r} - \boldsymbol{r}'|} \right]_{z' \longrightarrow -z'}.
    \label{greenFunctionNeummanEq}
\end{equation}
This expression results in:
\[
\lim_{z' \to 0} G_N(\boldsymbol{r},\boldsymbol{r}') = -\frac{1}{2\pi} \frac{\delta(t' - t_{\boldsymbol{r}})}{|\boldsymbol{r} - \boldsymbol{r}'|} \Big|_{z' = 0},
\]
when evaluated near the surface.

This Green's function {\color{black}satisfies the Neumann} boundary conditions on $\partial \mathcal{D}$, as demonstrated in Appendix \ref{BoundaryConditionSectionLabel}. Therefore, {\color{black}we have:}
\[
\boldsymbol{\Theta}(\boldsymbol{r},t) = \frac{1}{2\pi} \int_\mathbb{T} \int_{\mathcal{G} \in \mathbb{R}^2} \left. \frac{\delta(t' - t_{\boldsymbol{r}})}{|\boldsymbol{r} - \boldsymbol{r}'|} \right|_{z' = 0} \boldsymbol{\mathcal{W}}(u',\phi',t') u' \, du' \, d\phi' \, dt'.
\]
Including the retarded time definition, the electric vector potential {\color{black}becomes:}
\begin{equation}
\boxed{
\boldsymbol{\Theta}(\boldsymbol{r},t) = \frac{1}{2\pi} \int_{\mathcal{G} \in \mathbb{R}^2} \frac{1}{|\boldsymbol{r} - \boldsymbol{r}'|} \boldsymbol{\mathcal{W}} \left( u', \phi', t - \frac{|\boldsymbol{r} - \boldsymbol{r}'|}{c} \right) u' \, du' \, d\phi',
}    
\label{electricVectoPotentialFinalEq}
\end{equation}
for which the time integral is suppressed. 

According to Eq.~(\ref{electricVectoPotentialFinalEq}), the radiated electric vector potential can be computed by integrating the vector $\boldsymbol{\mathcal{W}} / {|\boldsymbol{r} - \boldsymbol{r}'|}$ over the gap region $\mathcal{G}$ between the electrodes.

{\color{black}It is important to note} the sign of the argument of the weight vector $\boldsymbol{\mathcal{W}} \left( u', \phi', t \pm \frac{|\boldsymbol{r} - \boldsymbol{r}'|}{c} \right)$. {\color{black}If the sign is} negative, then Eq.~(\ref{greenFunctionNeummanEq}) {\color{black}describes two point} sources, and the plane $z = 0$ acts as a receptor of incoming waves. On the other hand, {\color{black}if} the sign is positive, Eq.~(\ref{greenFunctionNeummanEq}) deals with two {\color{black}point sinks, adapting} the Green's function to a scenario where electrodes on the $z = 0$ plane radiate waves.

\section{Electric Field}

In this section, we integrate the expression {\color{black}in Eq.~}(\ref{electricVectoPotentialFinalEq}) to calculate the radiated electric field. The integral expressions establish a direct connection between the electric field and wire and ribbon loop antennas. Consequently, traditional methods applicable to such problems can be readily employed for Planar Blade antennas. Towards the end of the section, we discuss the definition of the weight vector and the principles of charge conservation.

\subsection{General expression for the dipole planar blade}

Now that we have the time-dependent electric vector potential for the dipole planar blade radiator, we can use the result from Eq.~(\ref{electricVectoPotentialFinalEq}) to compute the electric field as follows:
\[
\boldsymbol{E}(\boldsymbol{r},t) = \nabla \times \boldsymbol{\Theta}(\boldsymbol{r},t) = \frac{1}{2\pi} \int_{\mathcal{G} \in \mathbb{R}^2} \nabla \times \left[ \frac{1}{|\boldsymbol{r} - \boldsymbol{r}'|}  \boldsymbol{\mathcal{W}}\left(\boldsymbol{r}', t - \frac{|\boldsymbol{r} - \boldsymbol{r}'|}{c}\right) \right] d^2 \boldsymbol{r}'.
\]
We can further develop the term within the integral as:
\[
\nabla \times \left[ \frac{1}{|\boldsymbol{r} - \boldsymbol{r}'|}  \boldsymbol{\mathcal{W}}\left(\boldsymbol{r}', t - \frac{|\boldsymbol{r} - \boldsymbol{r}'|}{c}\right) \right] = \frac{1}{|\boldsymbol{r} - \boldsymbol{r}'|} \nabla \times \boldsymbol{\mathcal{W}}(\boldsymbol{r}', t_{\boldsymbol{r}}) + \left( \nabla \frac{1}{|\boldsymbol{r} - \boldsymbol{r}'|} \right) \times \boldsymbol{\mathcal{W}}(\boldsymbol{r}', t_{\boldsymbol{r}}),
\]
where we have employed the vector calculus identity $\nabla \times (f \boldsymbol{F}) = f \nabla \times \boldsymbol{F} + (\nabla f) \times \boldsymbol{F}$, {\color{black}with $f$ being} a scalar function and $\boldsymbol{F}$ a vector field. The gradient term can be computed as:
\[
\nabla \frac{1}{|\boldsymbol{r} - \boldsymbol{r}'|} = -\frac{\boldsymbol{r} - \boldsymbol{r}'}{|\boldsymbol{r} - \boldsymbol{r}'|^3}.
\]
The term $\nabla \times \boldsymbol{\mathcal{W}}(\boldsymbol{r}', t_{\boldsymbol{r}})$ can be simplified by developing each spatial coordinate term. {\color{black}For example, the $x$-component is:}
\begin{align*}
\left[ \nabla \times \boldsymbol{\mathcal{W}}(\boldsymbol{r}', t_{\boldsymbol{r}}) \right]_x &= \frac{\partial}{\partial y} \mathcal{W}_z(\boldsymbol{r}', t_{\boldsymbol{r}}) - \frac{\partial}{\partial z} \mathcal{W}_z(\boldsymbol{r}', t_{\boldsymbol{r}})  \\
&= \frac{\partial}{\partial t_{\boldsymbol{r}}} \mathcal{W}_z(\boldsymbol{r}', t_{\boldsymbol{r}}) \left( \frac{\partial t_{\boldsymbol{r}}}{\partial y} - \frac{\partial t_{\boldsymbol{r}}}{\partial z} \right) \\
&= \frac{1}{c|\boldsymbol{r} - \boldsymbol{r}'|} \frac{\partial}{\partial t_{\boldsymbol{r}}} \mathcal{W}_z(\boldsymbol{r}', t_{\boldsymbol{r}}) \left[ -(y - y') + (z - z') \right] \\
&= \frac{1}{c|\boldsymbol{r} - \boldsymbol{r}'|} \left[ \frac{\partial}{\partial t_{\boldsymbol{r}}} \mathcal{W}(\boldsymbol{r}', t_{\boldsymbol{r}}) \times (\boldsymbol{r} - \boldsymbol{r}') \right]_x.
\end{align*}
Similarly, the other components can be calculated, and the vector form result is:
\[
\nabla \times \boldsymbol{\mathcal{W}}(\boldsymbol{r}', t_{\boldsymbol{r}}) = \frac{1}{c} \frac{\partial}{\partial t_{\boldsymbol{r}}} \boldsymbol{\mathcal{W}}(\boldsymbol{r}', t_{\boldsymbol{r}}) \times \frac{\boldsymbol{r} - \boldsymbol{r}'}{|\boldsymbol{r} - \boldsymbol{r}'|}.
\]
Therefore, the term within the integral takes the following form:
\[
\nabla \times \left[ \frac{1}{|\boldsymbol{r} - \boldsymbol{r}'|}  \boldsymbol{\mathcal{W}}\left(\boldsymbol{r}', t - \frac{|\boldsymbol{r} - \boldsymbol{r}'|}{c}\right) \right] = \frac{\partial}{\partial t_{\boldsymbol{r}}} \boldsymbol{\mathcal{W}}(\boldsymbol{r}', t_{\boldsymbol{r}}) \times \frac{1}{c} \frac{\boldsymbol{r} - \boldsymbol{r}'}{|\boldsymbol{r} - \boldsymbol{r}'|^2} + \boldsymbol{\mathcal{W}}(\boldsymbol{r}', t_{\boldsymbol{r}}) \times \frac{\boldsymbol{r} - \boldsymbol{r}'}{|\boldsymbol{r} - \boldsymbol{r}'|^3}.
\]
This allows us to compute the electric field from:
\begin{equation}
\boxed{
    \boldsymbol{E}(\boldsymbol{r},t) = \frac{\mbox{sgn}(z)}{2\pi}\int_{\mathcal{G}\in\mathbb{R}^2} \left[ \frac{\boldsymbol{\mathcal{W}}(\boldsymbol{r}',t_{\boldsymbol{r}})}{|\boldsymbol{r}-\boldsymbol{r}'|^3} + \frac{\partial_{t_{\boldsymbol{r}}}\boldsymbol{\mathcal{W}}(\boldsymbol{r}',t_{\boldsymbol{r}})}{c|\boldsymbol{r}-\boldsymbol{r}'|^2} \right]\times(\boldsymbol{r}-\boldsymbol{r}') d^2\boldsymbol{r}'
    }.
\label{electricFieldJefimenkoEq}    
\end{equation}

\subsubsection{Analogy with wire and ribbon loop antenna radiation}
\label{analogySection}
\begin{figure}[H]
\centering
\includegraphics[width=0.50\textwidth]{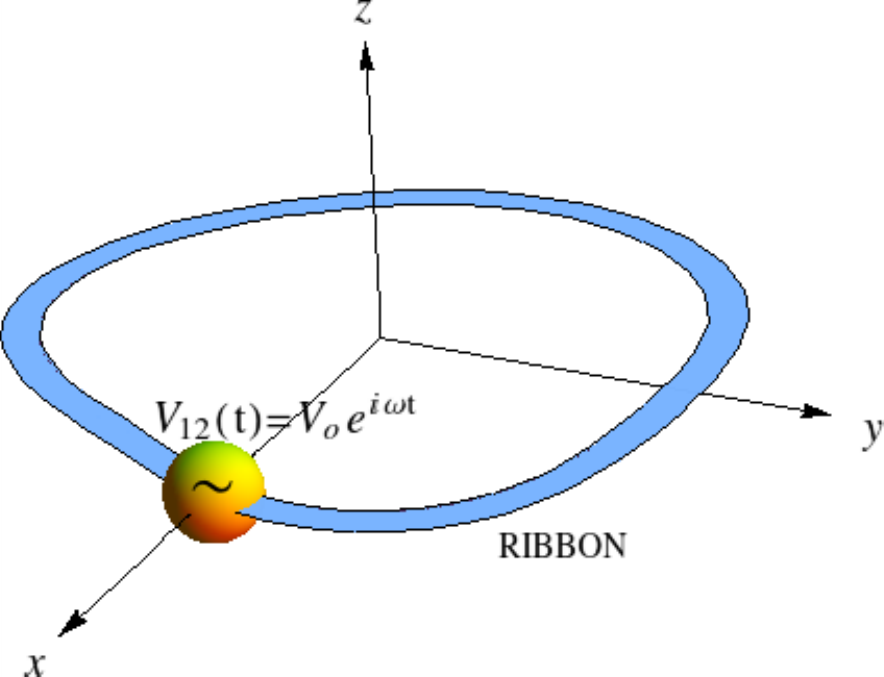}
\caption[The ribbon system.]{Ribbon fed by a time-dependent voltage.}
\label{ribbonFig}
\end{figure}

Let us consider the radiation problem produced by a {\color{black}surface current density} $\boldsymbol{K}(\boldsymbol{r},t)$ on a ribbon loop in the two-dimensional region $\mathcal{G}$, as shown in Fig.~\ref{ribbonFig}.
\label{SectionAnalogiesRibbonAntennas}

This magnetic vector potential is mathematically equivalent to the electric vector potential of the {\color{black}Planar Dipole Blade Antenna} (PDBA). The radiated electromagnetic field due to the ribbon is described by Jefimenko's equations\footnote{The corresponding Jefimenko equation for the electric field is given by:
\[
    \boldsymbol{E}_{ribbon}(\boldsymbol{r},t) = \frac{\mu_0}{4\pi\epsilon_0} \int_{\mathbb{R}^3} \left[ \frac{(\boldsymbol{r} - \boldsymbol{r'})}{|\boldsymbol{r} - \boldsymbol{r}'|^3} \rho(\boldsymbol{r}',t_{\boldsymbol{r}}) + \frac{(\boldsymbol{r} - \boldsymbol{r}')}{c|\boldsymbol{r} - \boldsymbol{r}'|^2} \partial_{t_{\boldsymbol{r}}} \rho(\boldsymbol{r}',t_{\boldsymbol{r}}) - \frac{\partial_{t_{\boldsymbol{r}}} \boldsymbol{J}(\boldsymbol{r}',t_{\boldsymbol{r}})}{c^2 |\boldsymbol{r} - \boldsymbol{r}'|} \right] d^3 \boldsymbol{r}',
\]
where $\boldsymbol{J}(\boldsymbol{r},t)$ is the volume current density defined as:
\[
\boldsymbol{J}(\boldsymbol{r},t) = \boldsymbol{K}(\boldsymbol{r},t) \delta(z) \hspace{0.5cm} \text{if} \hspace{0.5cm} \boldsymbol{r} \in \mathcal{G}, \hspace{0.5cm} \text{otherwise} \hspace{0.5cm} 0.
\]

In free space, there is no volume charge density, meaning that $\rho(\boldsymbol{r},t) = 0$ for $\boldsymbol{r} \notin \mathcal{G}$. For generality, we can express $\rho(\boldsymbol{r},t) = \sigma(\boldsymbol{r},t) \delta(z)$ for $\boldsymbol{r} \in \mathcal{G}$, with $\sigma$ representing the surface charge density. We assume that the ribbon is electrically neutral at all times, implying $\sigma = 0$. 
} 
\cite{griffiths2005introduction,jackson1999classical}.

The electric field of the ribbon loop, $\boldsymbol{E}_{ribbon}(\boldsymbol{r},t)$, is given by:
\begin{equation}
\boldsymbol{E}_{ribbon}(\boldsymbol{r},t) = -\frac{\mu_0}{4\pi} \int_{\mathcal{G} \in \mathbb{R}^2} \frac{\partial_{t_{\boldsymbol{r}}} \boldsymbol{K}(\boldsymbol{r}',t_{\boldsymbol{r}})}{|\boldsymbol{r} - \boldsymbol{r}'|} d^2 \boldsymbol{r}',
\label{ERibbonEq}    
\end{equation}
and its magnetic field, $\boldsymbol{B}_{ribbon}(\boldsymbol{r},t)$, is given by:
\begin{equation}
\boldsymbol{B}_{ribbon}(\boldsymbol{r},t) = \frac{\mu_0}{4\pi} \int_{\mathcal{G} \in \mathbb{R}^2} \left[ \frac{\boldsymbol{K}(\boldsymbol{r}',t_{\boldsymbol{r}})}{|\boldsymbol{r} - \boldsymbol{r}'|^3} + \frac{\partial_{t_{\boldsymbol{r}}} \boldsymbol{K}(\boldsymbol{r}',t_{\boldsymbol{r}})}{c|\boldsymbol{r} - \boldsymbol{r}'|^2} \right] \times (\boldsymbol{r} - \boldsymbol{r}') d^2 \boldsymbol{r}'.
\label{BRibbonEq}    
\end{equation}

The radiated magnetic field for this problem is obtained through the standard retarded magnetic vector potential $\boldsymbol{A}_{ribbon}$, {\color{black}with the relation} $\boldsymbol{B}_{ribbon}(\boldsymbol{r},t) = \nabla \times \boldsymbol{A}_{ribbon}(\boldsymbol{r},t)$. Thus, we have:
\begin{equation}
\boldsymbol{A}_{ribbon}(\boldsymbol{r},t) = \frac{\mu_0}{4\pi} \int_{\mathcal{G} \in \mathbb{R}^2} \frac{1}{|\boldsymbol{r} - \boldsymbol{r}'|} \boldsymbol{K} \left( \boldsymbol{r}', t - \frac{|\boldsymbol{r} - \boldsymbol{r}'|}{c} \right) d^2 \boldsymbol{r}.
\label{ARibbonEq}
\end{equation}

These expressions are \textit{mathematically equivalent} to the electric field in Eq.~(\ref{electricVectoPotentialFinalEq}) for $z > 0$. The weight vector $\boldsymbol{\mathcal{W}}(\boldsymbol{r},t)$ between electrodes in Eq.~(\ref{weightVectorEq}) \textit{plays the role} of the {\color{black}surface current density} $\boldsymbol{K}(\boldsymbol{r},t)$ on the ribbon.

\subsection{Divergence-less condition on $\boldsymbol{\mathcal{W}}$ and its relationship with the charge conservation law}

Having established an analogy between the weight vector $\boldsymbol{\mathcal{W}}$ for the PDBA and the surface {\color{black}current density} $\boldsymbol{K}$ for a ribbon antenna, we further explore the physical meaning of this relationship. This analogy holds {\color{black}for} two different radiating systems that share common wave fronts within the region $\mathcal{D}$.

The surface charge density on the electrode layers can be determined using Gauss' law:
\[
\sigma(\boldsymbol{r},t) = 2 \epsilon \lim_{z \rightarrow 0^+} E_z(\boldsymbol{r},t), 
\]
where $\sigma$ represents the surface charge density. This expression arises from the application of Gauss' law in the limit as $z$ approaches zero, reflecting charge conservation within the system. In the gap region $\mathcal{G}$ between the electrodes, there is no charge density due to the specific definition of the gap. Therefore, for all points $\boldsymbol{r}$ within $\mathcal{G}$, {\color{black}we have} $\boldsymbol{E}(\boldsymbol{r},t) \cdot \hat{z} = 0$. This implies that the time-dependent electric field over the gap region {\color{black}generates} a magnetic field. According to Ampère-Maxwell's law, we have:
\begin{equation}
    \nabla \times \boldsymbol{B}(\boldsymbol{r},t) =  \mu_0 \epsilon_0 \frac{\partial \boldsymbol{E}(\boldsymbol{r},t)}{\partial t} \implies \oint_{\partial S} \boldsymbol{B}(\boldsymbol{r},t) \cdot d\boldsymbol{r} = \mu_0 \epsilon_0 \frac{d}{dt} \int_S \boldsymbol{E}(\boldsymbol{r},t) \cdot d\boldsymbol{S}.
    \label{eq:simplifiedintegral}
\end{equation}

For simplicity, we demonstrate the integration over circular surface electrodes {\color{black}with $\mathcal{R}(\phi) = R$ and} a thickness of $\nu = \Delta R$. The surface $S$ for this simple case is defined as the region with boundaries in cylindrical coordinates:
\[
\partial S(u,\beta) = \mbox{arc}_{-}(u,\beta) \cup l_{+} \cup \mbox{arc}_{+}(u,\beta) \cup l_{-},
\]
where $\beta \in [0, 2\pi)$ represents a specified angle, and $u$ is a constant radial value within $(R - \delta R/2, R + \delta R/2)$. The boundaries consist of an arc path:
\[
\mbox{arc}_{-}(\beta) = \left\{ (u, \phi, -\delta h/2) : 0 < \phi < \beta \right\}, \quad \mbox{arc}_{+}(\beta) = \left\{ (u, \beta - \phi, \delta h/2) : 0 < \phi < \beta \right\},
\]
and vertical lines:
\[
l_{+}(u,\beta) = \left\{ (u, \theta, z) : -\delta h/2 < z < \delta h/2 \right\}, \quad l_{-} = \left\{ (u, 0, z) : -\delta h/2 < z < \delta h/2 \right\}.
\]

\begin{figure}[H]
\centering
\includegraphics[width=0.65\textwidth]{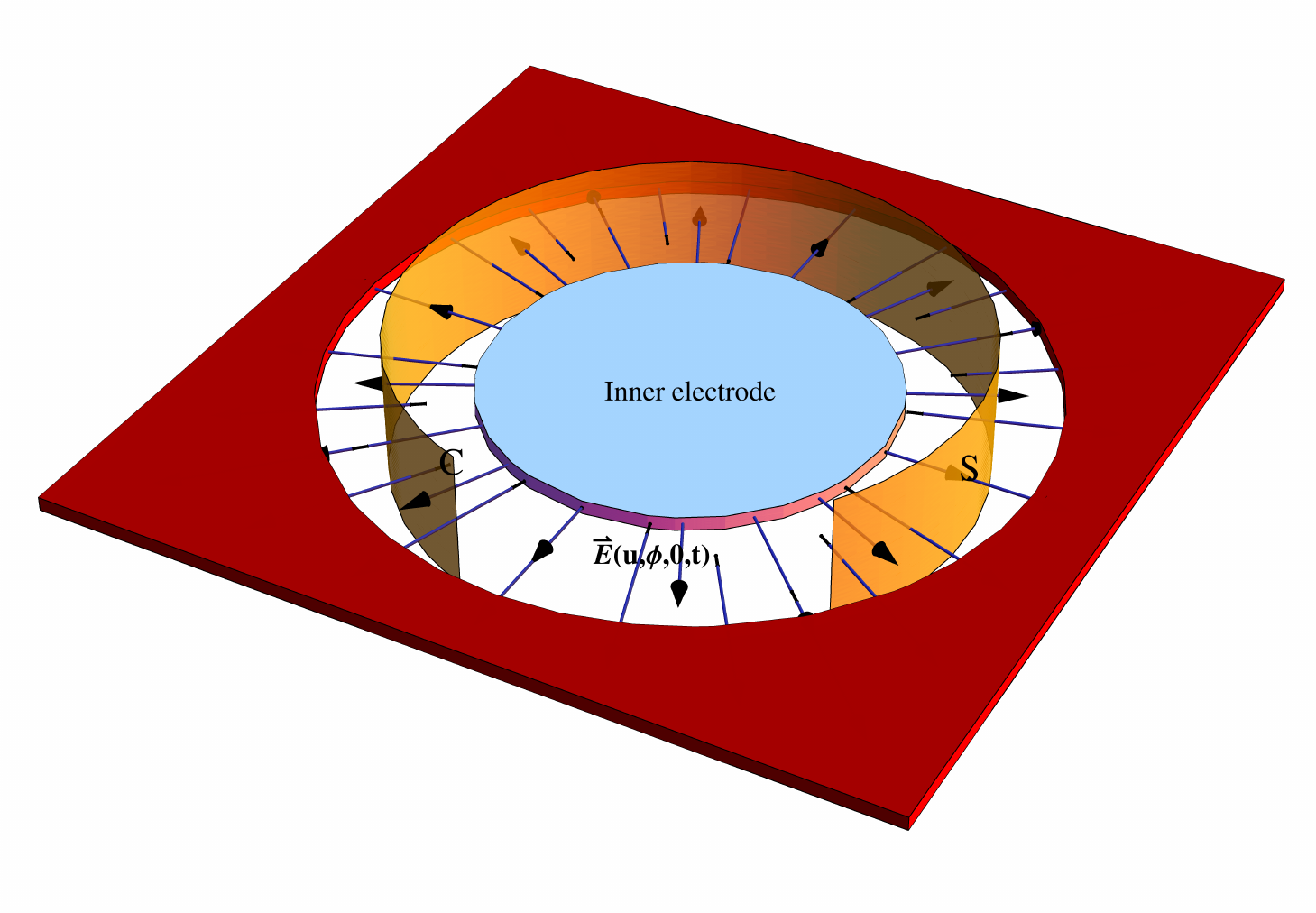}
\caption[Electric Vector Potential Field]{Electric vector potential field distribution across the gap region of the circular planar dipole blade radiator (isometric projection).}
\label{circularSE}
\end{figure}

If $\beta$ is chosen as $2\pi$ and $u = R$, the surface $S$ becomes a curved surface of radius $R$ and height $\delta h$, encompassing the mid-trajectory $c = \{(R, \phi, 0) : \phi \in [0, 2\pi)\}$ between the electrodes (see Fig.~\ref{circularSE}). Due to the symmetry of the system, the magnetic field circulation is expected to be negligible on $l_{\pm}$ as $\delta h$ tends to zero. {\color{black}Moreover,} the magnetic field should be uniform on the arc paths $\mbox{arc}_{\pm}(u,\beta)$, resulting in equal circulation.

Thus, {\color{black}the integral in Eq.~}(\ref{eq:simplifiedintegral}) can be expressed as:
\[
2 L(u,\beta) \delta h B(R,\delta h,t) = \mu_0 \epsilon_0 \frac{d}{dt} \int_S \boldsymbol{E}(\boldsymbol{r},t) \cdot d\boldsymbol{S},
\]
with $L(u,\beta) = \pi \beta u$ denoting the arc length of $\mbox{arc}_{\pm}(u,\beta)$. The flux of the electric field is given by:
\[
\int_S \boldsymbol{E}(\boldsymbol{r},t) \cdot d\boldsymbol{S} = \langle E(R,t) \rangle L(\beta) \delta h,
\]
where $\langle E(u,t) \rangle$ represents the spatial average of the electric field on $S$. Consequently, the magnetic field in the vicinity of the gap $\mathcal{G}$ can be expressed as:
\[
\boldsymbol{B}(u,z,t) = \frac{\mu_0 \epsilon_0}{2} \langle \dot{E}(u,t) \rangle \text{sgn}(z) \hat{\phi}
\]
for $z/\delta R \ll 1$. Since $\text{sgn}(0) = 0$ and $u$ is an arbitrary value, we can expect that:
\[
\lim_{z \to 0} \boldsymbol{B}(u,\phi,z,t) = 0 \hspace{0.5cm} \forall \hspace{0.5cm} (u,\phi) \in \mathcal{G}.
\]
That implies the vector potential in the gap is $\boldsymbol{A}_{\mathcal{G}} = \nabla \lambda$, where $\lambda$ is an arbitrary scalar function, and consequently, $\boldsymbol{B} = \nabla \times \boldsymbol{A}_{\mathcal{G}} = 0$ in the gap. The divergence of the weight vector $\boldsymbol{\mathcal{W}}$ within the gap region $\mathcal{G}$ can be computed as follows:
\[
\nabla \cdot \boldsymbol{\mathcal{W}}(\boldsymbol{r},t) = \frac{1}{u} \frac{\partial}{\partial u} (u \mathcal{W}_u) + \frac{1}{u} \frac{\partial \mathcal{W}_\phi}{\partial \phi}.
\]
Following similar procedures as with the curl terms, we evaluate $\nabla \cdot \boldsymbol{\mathcal{W}}(\boldsymbol{r},t)$ using Eq.~(\ref{weightVectorEq}), resulting in:
\begin{align*}
    \nabla \cdot \boldsymbol{\mathcal{W}}(u,\phi,t) &= \frac{1}{u} \frac{\partial}{\partial u} \left( \frac{\partial \Phi}{\partial u} - u \frac{\partial A_{\phi}}{\partial t} \right) + \frac{1}{u} \frac{\partial}{\partial \phi} \left( -\frac{\partial \Phi}{\partial u} + \frac{\partial A_u}{\partial t} \right) \\
    &= \frac{1}{u} \frac{\partial^2 \Phi}{\partial u \partial \phi} - \frac{1}{u} \frac{\partial^2}{\partial u \partial t} (u A_{\phi}) - \frac{1}{u} \frac{\partial^2 \Phi}{\partial \phi \partial u} + \frac{1}{u} \frac{\partial^2}{\partial \phi \partial t} \left( A_u \right) \\
    &= -\frac{1}{u} \frac{\partial}{\partial t} \left[ \frac{\partial}{\partial u} (u A_\phi) - \frac{\partial A_u}{\partial \phi} \right] \\
    &= - \frac{\partial}{\partial t} \left( \nabla \times \boldsymbol{A} \right)_z \\
    &= - \frac{\partial}{\partial t} B_z(\boldsymbol{r},t) = \frac{\partial}{\partial t} \left[ \boldsymbol{B}(\boldsymbol{r},t) \cdot \hat{n} \right] = 0 \hspace{0.5cm} \forall \hspace{0.5cm} \boldsymbol{r} \in \mathcal{G}.
\end{align*}
Thus, the divergence of the weight vector is {\color{black}simply the negative of} the perpendicular component of $\partial_t \boldsymbol{B}$ in the gap region. However, {\color{black}as seen in the case of} the circular planar blade radiator, the magnetic field near the gap has only the $\phi$-component, and it must be zero in the gap due to the reflection symmetry of the system with respect to the plane $z=0$. This demonstrates that the weight vector is divergence-less:
\begin{equation}
\boxed{
\nabla \cdot \boldsymbol{\mathcal{W}}(\boldsymbol{r},t) = 0 \hspace{0.5cm} \forall \hspace{0.5cm} \boldsymbol{r} \in \mathcal{G}
}\hspace{0.5cm} \text{(Continuity condition)}.
\label{conservationAnalogyEq}
\end{equation}
This implies
\[
\frac{\partial}{\partial u} [u A_\phi(\boldsymbol{r},t)] = \frac{\partial A_u(\boldsymbol{r},t)}{\partial \phi} \hspace{0.5cm} \forall \hspace{0.5cm} \boldsymbol{r} \in \mathcal{G}.
\]

In section~\ref{analogySection}, we include some comments about the analogy between the planar dipole blade and the radiation due to a ribbon antenna. Under this analogy, the weight vector $\boldsymbol{\mathcal{W}}(\boldsymbol{r},t)$ plays {\color{black}the role of a surface current density due to} the mathematical equivalence between Eqs.~(\ref{electricFieldJefimenkoEq}) and Eqs.~(\ref{BRibbonEq}) in the real half-space $\mathcal{D}$.

{\color{black}For the case of a wire antenna, we} know that charge conservation is {\color{black}expressed by:}
\[
\nabla \cdot \boldsymbol{J}(\boldsymbol{r},t) + \frac{\partial \rho}{\partial t} = 0,
\]
assuming the wire is electrically neutral{\color{black}, implying} $\partial_t \rho = 0$. In two dimensions, {\color{black}this reduces to the condition} $\nabla \cdot \boldsymbol{K}(\boldsymbol{r},t) = 0$ along the ribbon. {\color{black}Thus,} Eq.~(\ref{conservationAnalogyEq}) is analogous to the charge conservation law of a neutral ribbon antenna over the gap region between electrodes.

\section{Harmonic radiation}

Having characterized the time-dependent electric and magnetic fields in the dipole blade system using the electric vector potential representation, we now introduce a harmonic pulse as the time-dependent potential applied to the electrodes. Specifically, we set the scalar potential of a single electrode, denoted as $\Phi_o(\boldsymbol{r},t)$, to $\Phi_o(t) = V_o \exp(\boldsymbol{i}\omega t)$. For simplicity, we focus on the gapless limit and proceed to analyze the resulting low- and high-frequency radiation spectra.

\subsection{Radiated electromagnetic fields}

We consider the application of a harmonic pulse to the dipole blade system, with the scalar potential of a single electrode defined as $\Phi_o(t) = V_o \exp(\boldsymbol{i}\omega t)$. In this context, the electric potential in the entire domain $\mathcal{D}$, including the gap region $\mathcal{G}$, becomes a harmonic function: $\Phi(\boldsymbol{r},t) = \Phi_o(\boldsymbol{r}) \exp(\boldsymbol{i}\omega t)$. 

Following the Lorentz gauge condition, we derive the divergence of the magnetic vector potential $\boldsymbol{A}(\boldsymbol{r},t)$:
\[
\nabla \cdot \boldsymbol{A}(\boldsymbol{r},t) + \frac{1}{c^2}\frac{\partial \Phi(\boldsymbol{r},t)}{\partial t} = 0 \hspace{0.5cm} \implies \hspace{0.5cm} \nabla \cdot \boldsymbol{A}(\boldsymbol{r},t) = -\frac{\boldsymbol{i}\omega}{c^2} \Phi_o(\boldsymbol{r}) \exp(\boldsymbol{i}\omega t).
\]
This leads to the representation of $\boldsymbol{A}(\boldsymbol{r},t)$ in the harmonic form $\boldsymbol{A}(\boldsymbol{r},t) = \boldsymbol{A}_o(\boldsymbol{r}) \exp(\boldsymbol{i}\omega t)$. Thus, the weight vector defined in Eq.~(\ref{weightVectorEq}) takes the form:
\begin{align*}
\boldsymbol{\mathcal{W}}(u,\phi,t) &= \left[\left(\frac{1}{u}\frac{\partial \Phi_o}{\partial \phi} - \boldsymbol{i}\omega(A_o)_\phi\right)\hat{u} + \left(-\frac{\partial \Phi_o}{\partial u} + \boldsymbol{i}\omega(A_o)_u\right)\hat{\phi}\right]\exp(\boldsymbol{i}\omega t) \\
&= \boldsymbol{\mathcal{W}}_o(u,\phi)\exp(\boldsymbol{i}\omega t).
\end{align*}
By combining this expression with Eq.~(\ref{electricVectoPotentialFinalEq}), we determine the electric vector potential:
\[
\boldsymbol{\Theta}(\boldsymbol{r},t) = \boldsymbol{\Theta}_o(\boldsymbol{r}) \exp(\boldsymbol{i}\omega t),
\]
where
\begin{equation}
\boldsymbol{\Theta}_o(\boldsymbol{r}) = \frac{1}{2\pi} \int_{\mathcal{G} \in \mathbb{R}^2} \frac{1}{|\boldsymbol{r} - \boldsymbol{r}'|} \boldsymbol{\mathcal{W}}_o\left(u',\phi'\right) \exp\left(-\boldsymbol{i}\omega \frac{|\boldsymbol{r} - \boldsymbol{r}'|}{c}\right) u'du'd\phi'.    
\label{ThetaoEq}
\end{equation}
Now, the Lorentz gauge-like condition in Eq.~(\ref{lorenztGaugeLikeConditionEq}) becomes:
\[
\nabla \cdot \boldsymbol{\Theta}(\boldsymbol{r},t) = \mu_0 \frac{\partial \Psi(\boldsymbol{r},t)}{\partial t} = \boldsymbol{i}\mu_0\omega \Psi_o(\boldsymbol{r}) \exp(\boldsymbol{i}\omega t) 
\]
for harmonic radiation. We also define the scalar magnetic potential $\Psi(\boldsymbol{r},t)$ as:
\begin{equation}
\Psi(\boldsymbol{r},t) = \frac{1}{\boldsymbol{i}\mu_0\omega} \nabla \cdot \boldsymbol{\Theta}(\boldsymbol{r},t) \hspace{1.0cm} \text{(scalar magnetic potential)}.  
\label{scalarMagneticPotentialHarmonicEq}
\end{equation}
This satisfies the Lorentz gauge-like condition.

The resulting electromagnetic fields for $z > 0$ are then computed using Eqs.~(\ref{electricFieldIdentityEq}) and (\ref{magneticFieldIdentityEq}):

\begin{align}
\boldsymbol{B}(\boldsymbol{r},t) &= \frac{\boldsymbol{i}}{\omega} \nabla\left[\nabla \cdot \boldsymbol{\Theta}(\boldsymbol{r},t)\right]  + \frac{1}{c^2} \frac{\partial \boldsymbol{\Theta}(\boldsymbol{r},t)}{\partial t} & \hspace{1.0cm} \text{(Magnetic field)}, \label{BFieldHarmonicEq}\\
\boldsymbol{E}(\boldsymbol{r},t) &= \nabla \times \boldsymbol{\Theta}(\boldsymbol{r},t) & \hspace{1.0cm} \text{(Electric field)}.
\label{EFieldHarmonicEq}
\end{align}

Thus, for harmonic radiation in the planar dipole blade system, the computation of the electric vector potential $\boldsymbol{\Theta}(\boldsymbol{r},t)$ is sufficient. Other quantities, such as the scalar {\color{black}magnetic} potential in Eq.~(\ref{scalarMagneticPotentialHarmonicEq}) and the electromagnetic fields given by Eqs.~(\ref{BFieldHarmonicEq}) and (\ref{EFieldHarmonicEq}), can be derived from the specific expression of $\boldsymbol{\Theta}(\boldsymbol{r},t)$.

\subsection{The gapless limit}

In the case of harmonic radiation with the inner and outer metallic layers being infinitely close to each other, the gap thickness tends to zero ($\nu \rightarrow 0$). In this limit, the scalar potential $\Phi(u,\phi,0)$ on the $z=0$ plane can be expressed as:
\[
\lim_{\nu \to 0} \Phi(u,\phi,0) = V_o \left[1 - H\left(|(u,\phi) - (\mathscr{R}(\phi),\phi)|\right)\right] \exp(\boldsymbol{i}\omega t).
\]
Here, $(\mathscr{R}(\phi),\phi)$ represents the parametric representation of the contour $\mathcal{A}_{in}$, and $H(z)$ is the Heaviside step function (refer to Fig.~\ref{theSystemFig}). In the gapless limit, the components of the weight vector become:
\begin{align*}
\lim_{\nu \to 0}\mathscr{W}_u &= \frac{1}{u} \left[ - V_o \frac{\partial}{\partial\phi} H(u - \mathscr{R}(\phi)) \right] \exp(\boldsymbol{i}\omega t) - \partial_t A_\phi(\mathscr{R}(\phi),\phi,0) \\
&= \left[ \frac{\dot{\mathscr{R}}(\phi)}{u} \delta(u - \mathscr{R}(\phi)) V_o - \boldsymbol{i}\omega A_\phi(\mathscr{R}(\phi),\phi,0) \right] \exp(\boldsymbol{i}\omega t),
\end{align*}
and
\begin{align*}
\lim_{\nu \to 0} \mathscr{W}_{\phi} &= - \left[ - V_o \frac{\partial}{\partial u} H(u - \mathscr{R}(\phi)) \right] \exp(\boldsymbol{i}\omega t) + \partial_t A_u(\mathscr{R}(\phi),\phi,0) \\
&= \left[ \delta(u - \mathscr{R}(\phi)) V_o + \boldsymbol{i}\omega A_u(\mathscr{R}(\phi),\phi,0) \right] \exp(\boldsymbol{i}\omega t).
\end{align*}
Here, $\delta(z)$ represents the Dirac delta function. It is essential to ensure that the magnetic field on the mid-trajectory $c_{mid}$ is zero ($\boldsymbol{B} = \nabla \times \boldsymbol{A} = 0$). Additionally, the vector potential must satisfy the Lorentz-Gauge condition for harmonic radiation:
\[
\nabla \cdot \boldsymbol{A}_o(\boldsymbol{r}) = \frac{\boldsymbol{i}\omega}{c^2} \Phi_o(\boldsymbol{r}).
\]

\subsubsection{Low frequency radiation}

In the case of low-frequency radiation, where the $\omega/c^2$ term is small, the scalar potential on the gap region is such that $0 < \Phi_o(u,\phi,0) < V_o$. Consequently, the divergence $\nabla \cdot \boldsymbol{A}_o(\mathcal{R}(\phi),\phi,0)$ is approximately zero, given the Coulomb gauge-like condition. This approximation allows us to simplify the magnetic vector potential on $c_{mid}$ as follows:
\[
\lim_{\nu \to 0} \boldsymbol{\mathscr{W}}(\boldsymbol{r},t) = \delta(u - \mathscr{R}(\phi)) \left[ \frac{\dot{\mathscr{R}}(\phi)}{u} \hat{u}(\phi) + \hat{\phi}(\phi) \right] \exp(\boldsymbol{i}\omega t).
\]
The electric vector potential, in this case, becomes:
\begin{align*}
\lim_{\nu \to 0} \boldsymbol{\Theta}(\boldsymbol{r},t) &= \frac{V_o}{2\pi} \exp(\boldsymbol{i}\omega t) \int_{\mathscr{G}} \frac{u' du' d\phi'}{|\boldsymbol{r} - \boldsymbol{r}'|} \delta(u' - \mathscr{R}(\phi')) \left[ \frac{\dot{\mathscr{R}}(\phi')}{u'} \hat{u}(\phi') + \hat{\phi}(\phi') \right] \exp\left( - \boldsymbol{i}\omega \frac{|\boldsymbol{r} - \boldsymbol{r}'|}{c} \right) \\
&= \frac{V_o}{2\pi} \exp(\boldsymbol{i}\omega t) \int_0^{2\pi} \frac{\mathscr{R}(\phi') d\phi'}{|\boldsymbol{r} - (\mathscr{R}(\phi'),\phi')|} \left[ \frac{\dot{\mathscr{R}}(\phi')}{\mathscr{R}(\phi')} \hat{u}(\phi') + \hat{\phi}(\phi') \right] \exp\left( - \boldsymbol{i}\omega \frac{|\boldsymbol{r} - \boldsymbol{r}'|}{c} \right) \\
&= \frac{V_o}{2\pi} \exp(\boldsymbol{i}\omega t) \int_0^{2\pi} \frac{\left[ \dot{\mathscr{R}}(\phi') \hat{u}(\phi') + \mathscr{R}(\phi') \hat{\phi}(\phi') \right] d\phi'}{|\boldsymbol{r} - (\mathscr{R}(\phi'),\phi')|} \exp\left( - \boldsymbol{i}\omega \frac{|\boldsymbol{r} - \boldsymbol{r}'|}{c} \right).
\end{align*}
Here, $d\boldsymbol{r}' = \left[ \dot{\mathscr{R}}(\phi') \hat{u}(\phi') + \mathscr{R}(\phi') \hat{\phi}(\phi') \right] d\phi'$ represents the differential displacement of $c_{mid}$. This leads to:
\begin{equation}
\lim_{\nu \to 0} \boldsymbol{\Theta}(\boldsymbol{r},t) = \frac{V_o}{2\pi} \exp(\boldsymbol{i}\omega t) \rcirclerightint_{c_{mid}} \frac{d\boldsymbol{r}'}{|\boldsymbol{r} - \boldsymbol{r}'|} \exp\left( - \boldsymbol{i}\omega \frac{|\boldsymbol{r} - \boldsymbol{r}'|}{c} \right) = \frac{1}{2\pi} \rcirclerightint_{c_{mid}} \frac{d\boldsymbol{r}'}{|\boldsymbol{r} - \boldsymbol{r}'|} \left[ V_o \exp(\boldsymbol{i}\omega t) \right]_{t \rightarrow t_{\boldsymbol{r}}}.
\label{gaplessLowFreqEq}
\end{equation}
Here, $t_{\boldsymbol{r}}$ represents the retarded time. This result is valid for low-frequency antennas where $\omega/c$ is small, and the metal plates can still be considered equipotential surfaces with a constant voltage $V_o$ along $c_{mid}$.

\subsubsection{High frequency radiation}

The approximation provided in Eq.~(\ref{gaplessLowFreqEq}) is no longer valid as the frequency increases. At higher frequencies, the emitted wave can significantly affect the distribution of currents on the metal plates and, consequently, on $c_{mid}$. Thus, the electric vector potential should be expressed in a more general form:
\[
\lim_{\nu \to 0} \boldsymbol{\Theta}(\boldsymbol{r},t) = \frac{1}{2\pi} \rcirclerightint_{c_{mid}} \frac{d\boldsymbol{r}'}{|\boldsymbol{r} - \boldsymbol{r}'|} \left[\mathcal{V}_{12}(\boldsymbol{r}') \exp(\boldsymbol{i}\omega t)\right]_{t \rightarrow t_{\boldsymbol{r}}}.
\]
Here, $\mathcal{V}_{12}(\boldsymbol{r}')$ represents the potential difference between the plates along $c_{mid}$. In general, $\mathcal{V}_{12}(\boldsymbol{r}')$ depends on the emission frequency, the geometry of $c_{mid}$, and the position of the voltage feed $V_{12}(t)$. In practice, we need to solve the equation:
\begin{equation}
\boldsymbol{\Theta}_o^{gapless}(\boldsymbol{r}) = \frac{1}{2\pi} \rcirclerightint_{c_{mid}} \frac{d\boldsymbol{r}'}{|\boldsymbol{r} - \boldsymbol{r}'|} \mathcal{V}_{12}(\boldsymbol{r}') \exp(-\boldsymbol{i}\kappa |\boldsymbol{r} - \boldsymbol{r}'|).
\label{ThetaWireLikeEq}
\end{equation}
Here, $\mathcal{V}_{12}(\boldsymbol{r}')$ is an unknown function that depends on the specific frequency of emission, the geometry of $c_{mid}$, and the position of the voltage feed $V_{12}(t)$. This equation is analogous to the magnetic vector potential due to an infinitely thin wire loop antenna:
\[
\boldsymbol{A}_o^{wire} = \frac{\mu_0}{4\pi} \int_{c_{mid}} \frac{d\boldsymbol{r}'}{|\boldsymbol{r} - \boldsymbol{r}'|} I(\boldsymbol{r}') \exp(-\boldsymbol{i}\kappa |\boldsymbol{r} - \boldsymbol{r}'|).
\]
In this analogy, $I(\boldsymbol{r}')$ represents the {\color{black}current intensity in the wire loop. Just as the current $I(\boldsymbol{r}')$ in the wire loop dictates the magnetic field in wire antennas, the function $\mathcal{V}_{12}(\boldsymbol{r}')$ governs the electric vector potential in the high-frequency limit for the gapless planar dipole blade antenna.}

\section{Illustrative example: planar blade antenna with a thin annular gap}
In this section, we consider a simplified case involving a planar blade antenna with a circular, thin gap operating under low-frequency voltage. The antenna's configuration is depicted in Fig.~\ref{circularBladeFig}.

\subsection{Low frequency in the far limit region $\kappa r \gg 1$}
\label{LowFreqFarLimitRegionSeclbl}

Since the gap is infinitely thin, we can employ Eq.~(\ref{gaplessLowFreqEq}) or, equivalently, Eq.~(\ref{ThetaWireLikeEq}) with $\mathcal{V}_{12}(\boldsymbol{r}')$ set to the operating voltage $V_o$. Thus, we have
\[
\boldsymbol{\Theta}_o(\boldsymbol{r}) = \frac{\mbox{sgn}(z)}{2\pi} \rcirclerightint_{c_{circle}} \frac{d\boldsymbol{r}'}{|\boldsymbol{r} - \boldsymbol{r}'|} V_o \exp(-\boldsymbol{i}\kappa |\boldsymbol{r} - \boldsymbol{r}'|).
\]

Here, $c_{circle}$ represents a circular loop with radius $R$. Due to the axial symmetry of the problem at low frequency, we can further simplify to
\[
\boldsymbol{\Theta}_o(\boldsymbol{r}) = \mbox{sgn}(z) \frac{V_o R}{2\pi} \int_0^{2\pi} \frac{\cos(\phi - \phi')}{\mathcal{r}(r,\theta,\phi - \phi')} \exp(-\boldsymbol{i}\kappa \mathcal{r}(r,\theta,\phi - \phi')) d\phi' \hat{\phi}(\boldsymbol{r}),
\]
with 
\[
\mathcal{r} = |\boldsymbol{r} - \boldsymbol{r}'| = \sqrt{r^2 + R^2 - 2rR\sin\theta\cos(\phi - \phi')}
\]
denoting the position in spherical coordinates $(r, \theta, \phi)$.

\begin{figure}[H]
\centering
\includegraphics[width=0.85\textwidth]{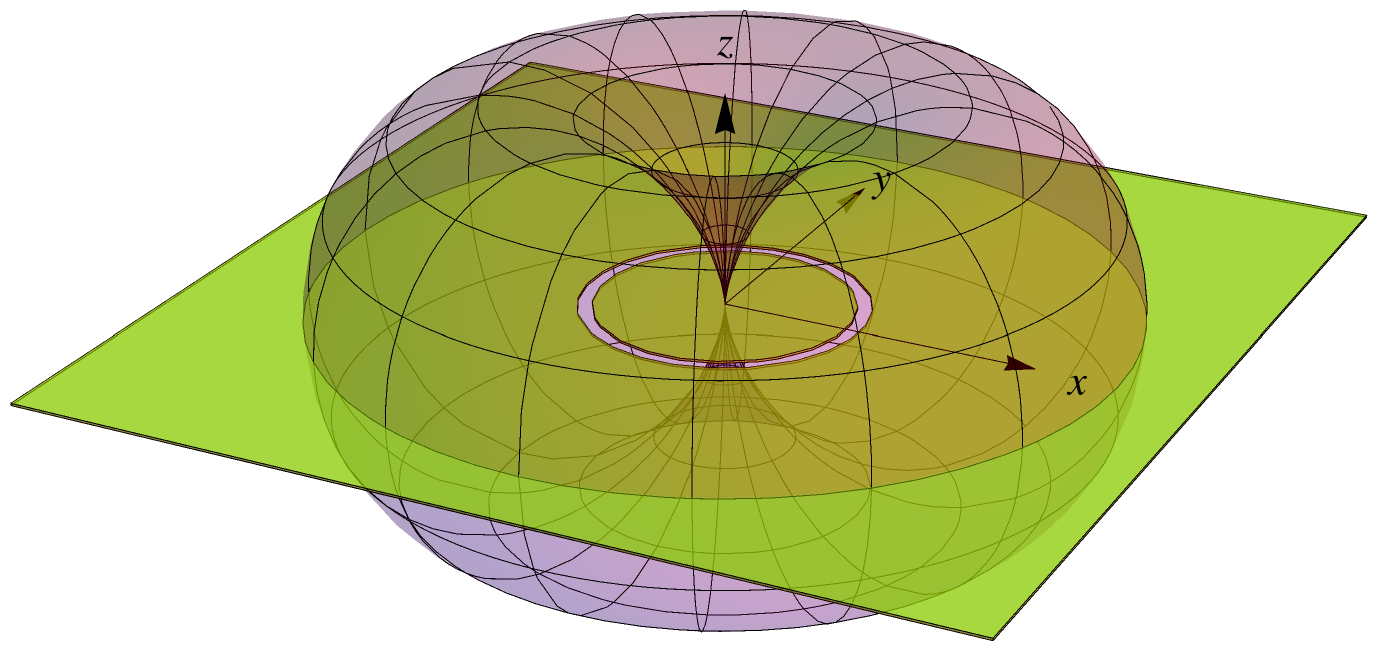}
\caption[The System]{A circular planar-blade antenna. The inner circular sheet has a radius $R$, and the outer sheet extends infinitely. These surfaces represent the radiation pattern at low frequencies.}
\label{circularBladeFig}
\end{figure} 

To simplify the problem, we focus on the far-field region where $r / R \gg 1$. In this regime, we can expand the integrand in a power series:
\[
f = \frac{1}{\mathcal{r}} \exp(-\boldsymbol{i}\kappa \mathcal{r}) = \sum_{m=0}^\infty \frac{R^m}{m!} \left[\frac{\partial^m f}{\partial R^m}\right]_{R=0}.
\]
Up to the second order, this expansion becomes:
\[
f = \left[1 + R \left(\frac{1}{r} + \boldsymbol{i}\frac{\omega}{c}\right) \sin\theta\cos(\phi - \phi')\right] \frac{\exp(-\boldsymbol{i}r\omega/c)}{r} + O(R^2).
\]
Consequently, the electric vector potential is given by:
\[
\boldsymbol{\Theta}_o(\boldsymbol{r}) = \mbox{sgn}(z) \frac{V_o R}{2\pi} \int_0^{2\pi} \left\{\left[1 + R \left(\frac{1}{r} + \boldsymbol{i}\frac{\omega}{c}\right) \sin\theta\cos\beta\right] \frac{\exp(-\boldsymbol{i}r\omega/c)}{r} d\beta \right\} \hat{\phi}(\boldsymbol{r}) + O(R^3),
\]
where $\beta = \phi - \phi'$. Computing the integral, we find:
\begin{equation}
\boldsymbol{\Theta}_o(\boldsymbol{r}) = \mbox{sgn}(z) \frac{V_o R^2}{2} \left(\frac{1}{r} + \boldsymbol{i}\frac{\omega}{c}\right) \sin\theta \frac{\exp(-\boldsymbol{i}r\omega/c)}{r} \hat{\phi}(\boldsymbol{r}),
\label{THETACircularApproxEq}
\end{equation}
and we have omitted terms of order $O(R^3)$. It is important to note that this result is only valid far from the antenna center.
\begin{figure}[H]
\centering
\includegraphics[width=0.5\textwidth]{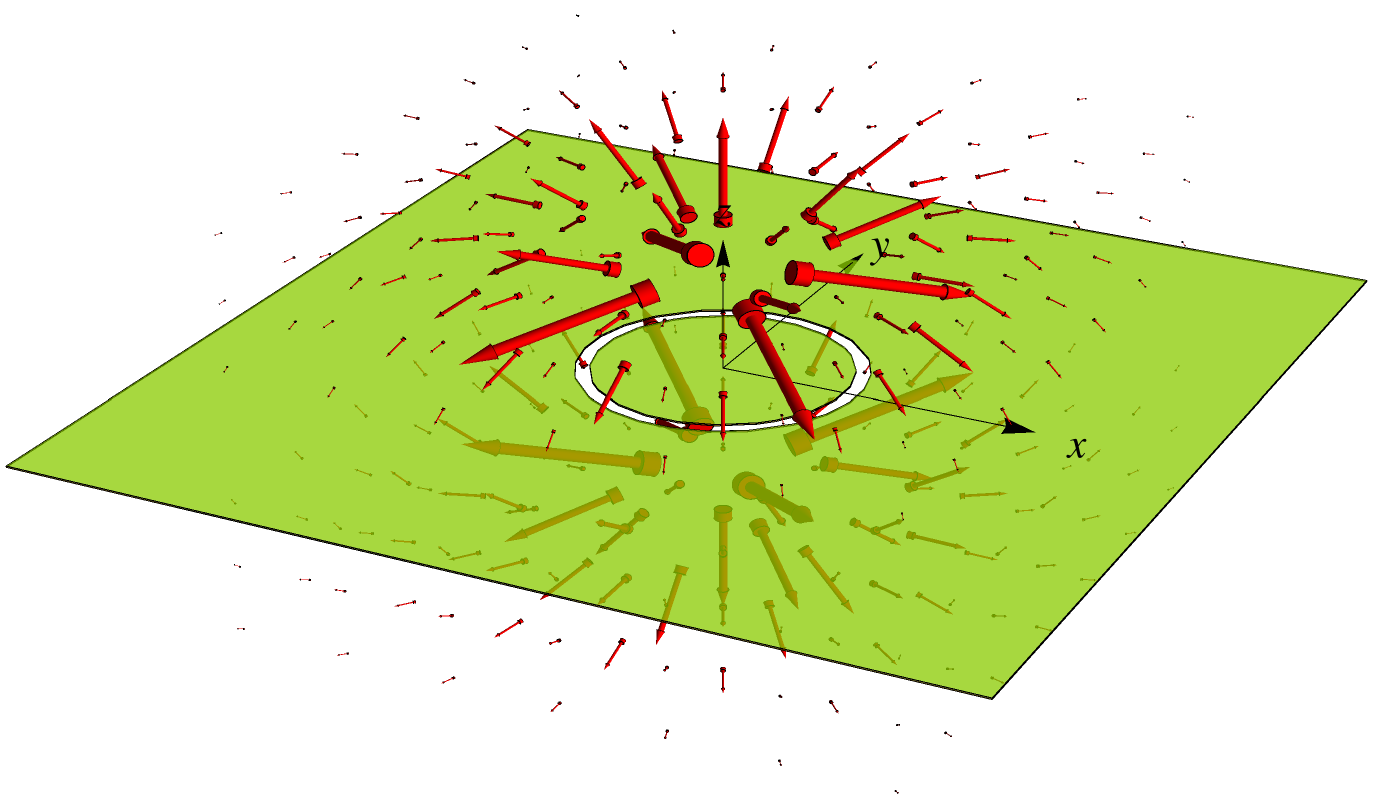}
\includegraphics[width=0.48\textwidth]{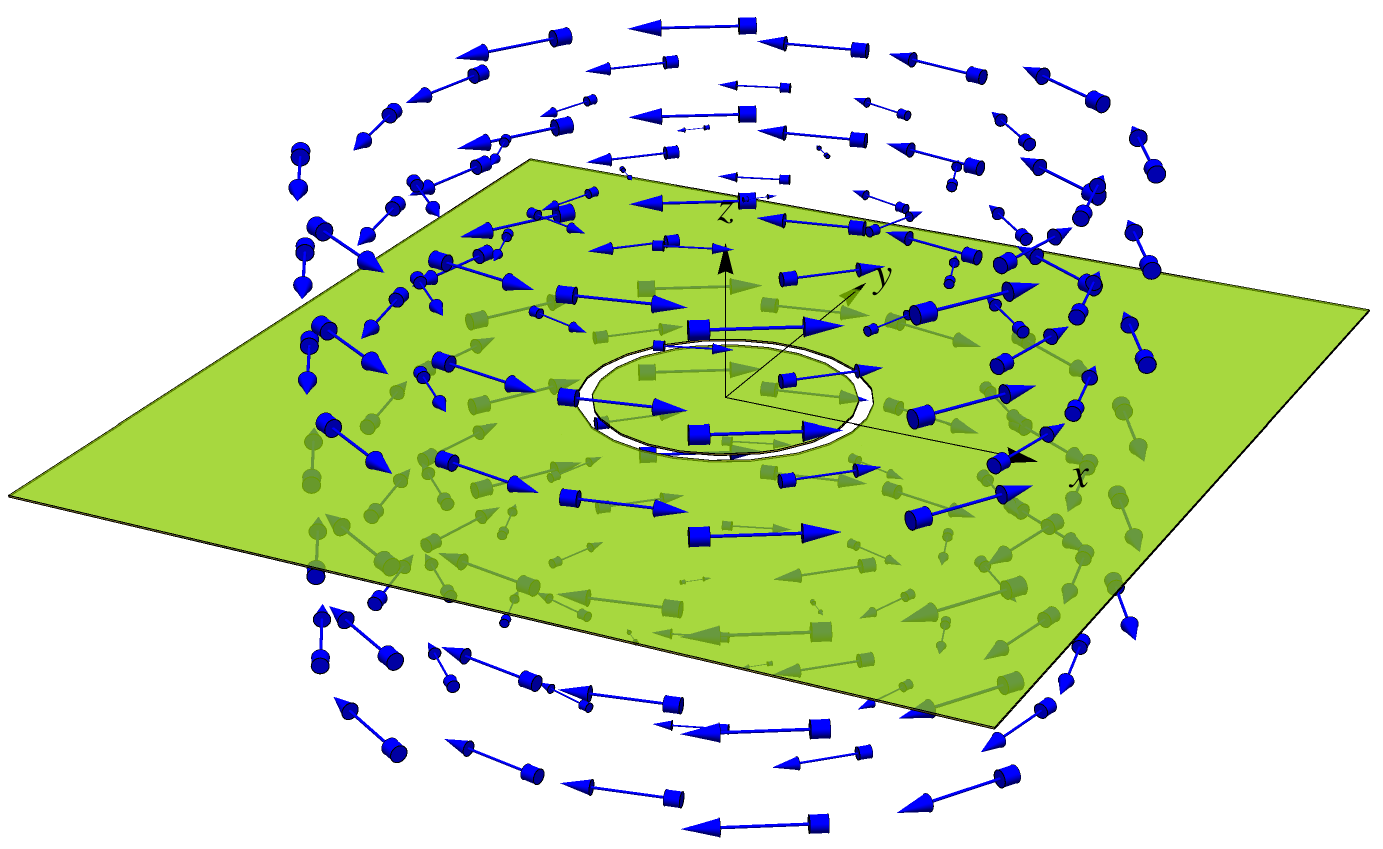}
\caption[The system.]{Radiated fields in the far region $\kappa r \gg 1$ in the gapless limit $\Delta R \rightarrow 0$. (Left) Electric fields given by Eqs.~(\ref{EorCircleEq})-(\ref{EophiCircleEq}), and (Right) magnetic fields according to Eq.~(\ref{BFieldCircleFarEq}).}
\label{fields3DFig}
\end{figure}
The electric field in the half-real plane, as derived from Eq.~(\ref{EFieldHarmonicEq}), is:
\[
\boldsymbol{E}_o = \nabla \times \boldsymbol{\Theta}_o = \frac{1}{r \sin\theta} \frac{\partial}{\partial \theta} \left(\Theta_o(r,\theta) \sin\theta\right) \hat{r} - \frac{1}{r} \frac{\partial}{\partial r} \left(r \Theta_o(r,\theta)\right) \hat{\theta} \hspace{0.5cm} z > 0.
\]
The components of the radiated electric field are:
\begin{align}
    E_{or}(r,\theta) &= \mbox{sgn}(z) \frac{V_o R^2}{r^2} \left(\frac{1}{r} + \boldsymbol{i} \frac{\omega}{c}\right) \cos\theta \exp(-\boldsymbol{i}r\omega/c), \label{EorCircleEq}\\
    E_{o\theta}(r,\theta) &= \mbox{sgn}(z) \frac{V_o R^2}{2r} \left(\frac{1}{r^2} + \boldsymbol{i} \frac{\omega}{cr} - \frac{\omega^2}{c^2}\right) \sin\theta \exp(-\boldsymbol{i}r\omega/c), \label{EothetaCircleEq}\\
    E_{o\phi} &= 0. \label{EophiCircleEq}
\end{align}
The magnetic scalar potential is defined as:
\[
\Psi(\boldsymbol{r},t) = \frac{1}{\boldsymbol{i}\mu_0\omega} \nabla \cdot \boldsymbol{\Theta}(\boldsymbol{r},t) = \frac{\exp{\left(\boldsymbol{i}\omega t\right)}}{\boldsymbol{i}\mu_0\omega} \nabla \cdot \left[\Theta_o(r,\theta)\hat{\phi}\right].  
\]
Taking the divergence of the electric vector potential in spherical coordinates, we find:
\[
\nabla \cdot \boldsymbol{\Theta}_o = \frac{1}{r^2} \frac{\partial}{\partial r} \left(r^2 \Theta_{or}\right) + \frac{1}{r \sin\theta} \frac{\partial}{\partial \theta} \left(\sin\theta \Theta_{o\theta}\right) + \frac{1}{r \sin\theta} \frac{\partial \Theta_{o\phi}}{\partial \phi}.
\]
{\color{black}Since,} according to Eq.~(\ref{THETACircularApproxEq}), the electric vector potential is along the $\phi$-component and not a function of the $\phi$ coordinate {\color{black}in the half-real region $z > 0$} ($\Theta_o = \Theta_o(r, \theta)$), then  
\[
\Psi(\boldsymbol{r},t) = \frac{\exp{\left(\boldsymbol{i}\omega t\right)}}{\boldsymbol{i}\mu_0\omega} \left[\frac{1}{r \sin\theta} \frac{\partial}{\partial \phi} \Theta_{o\phi}(r,\theta)\right] = 0.
\]
Hence, the magnetic scalar potential is null.

The magnetic field can be computed from Eq.~(\ref{EFieldHarmonicEq}) as:
\[
\boldsymbol{B}(\boldsymbol{r},t) = \frac{\boldsymbol{i}}{\omega} \nabla \left[\nabla \cdot \boldsymbol{\Theta}(\boldsymbol{r},t)\right]  + \frac{1}{c^2} \frac{\partial \boldsymbol{\Theta}(\boldsymbol{r},t)}{\partial t} = \boldsymbol{i} \frac{\omega}{c^2} \boldsymbol{\Theta}_o \exp{\left(\boldsymbol{i}\omega t\right)}.
\]
Once again, we used the fact that $\nabla \cdot \boldsymbol{\Theta}_o = 0$ according to Eq.~(\ref{THETACircularApproxEq}). The spatial part of the magnetic field is just proportional {\color{black}to the electric} vector potential, explicitly:
\begin{equation}
\boldsymbol{B}_o(\boldsymbol{r}) = \mbox{sgn}(z) \boldsymbol{i} \frac{\omega}{c^2} \frac{V_o R^2}{2} \left(\frac{1}{r} + \boldsymbol{i} \frac{\omega}{c}\right) \sin\theta \frac{\exp(-\boldsymbol{i}r\omega/c)}{r} \hat{\phi}(\boldsymbol{r}). 
\label{BFieldCircleFarEq}
\end{equation}

Now we can calculate the spatial component of the Poynting vector:
\[
\boldsymbol{S_o}(\boldsymbol{r}) = \frac{1}{2\mu_0} \left(\boldsymbol{E_o} \times \boldsymbol{B_o}^*\right).
\]
Explicitly, this is $\boldsymbol{S_o}(\boldsymbol{r}) = S_{or} \hat{r} + S_{o\theta} \hat{\theta}$, with $S_{or} = E_{o\theta} B^*_{o\phi}/(2\mu_0)$, $S_{o\theta} = -E_{or} B^*_{o\phi}/(2\mu_0)$, and $S_{o\phi} = 0$. The Poynting vector becomes:
\begin{align}
    S_{or}(r,\theta) &= \frac{1}{2\mu_0} \left(\frac{V_o R^2}{2r}\right)^2 \left(\frac{\omega}{c^2}\right) \sin^2\theta \left[\left(\frac{\omega}{c}\right)^3 + \frac{1}{\boldsymbol{i} r^3}\right], \label{SorCircleEq}\\
    S_{o\theta}(r,\theta) &= \frac{\boldsymbol{i}}{\mu_0} \frac{\left(V_o R^2\right)^2}{8r^3} \left(\frac{\omega}{c^2}\right) \sin(2\theta) \left[\left(\frac{\omega}{c}\right)^2 + \frac{1}{r^2}\right], \label{SothetaCircleEq}\\
    S_{o\phi} &= 0. \label{SophiCircleEq}
\end{align}

The time average of the radial component is:
\[
\langle S_r(\boldsymbol{r},t) \rangle = \mbox{Re} \left[\frac{1}{2\mu_0} \left(\boldsymbol{E_o} \times \boldsymbol{B_o}^*\right)_r\right] = \frac{1}{2\mu_0} \left(\frac{V_o R^2}{2r}\right)^2 \left(\frac{\omega}{c^2}\right) \sin^2\theta \left(\frac{\omega}{c}\right)^3,
\]
according to Eq.~(\ref{SorCircleEq}). Similarly,
\[
\langle S_\theta(\boldsymbol{r},t) \rangle = \mbox{Re} \left[\frac{1}{2\mu_0} \left(\boldsymbol{E_o} \times \boldsymbol{B_o}^*\right)_\theta\right] = 0
\]
since the result is purely imaginary.

The temporal average of the Poynting vector is radial and can be written as:
\[
\langle \boldsymbol{S}(\boldsymbol{r},t) \rangle = \frac{1}{2\eta_o} \left(\frac{V_o R^2}{2r}\right)^2 (\kappa R)^4 \sin^2\theta \hat{r},
\]
with 
\begin{align*}
\eta_o &= \sqrt{\frac{\mu_o}{\epsilon_o}} \hspace{0.5cm} \mbox{intrinsic impedance,}\\
\kappa &= \frac{\omega}{c} \hspace{0.5cm}\mbox{wave number.}    
\end{align*}
Thus, the radiation intensity becomes
\begin{equation}
U(\theta) = \langle S_r(\boldsymbol{r},t) \rangle r^2 = \frac{1}{2\eta_o} \left(\frac{V_o R^2}{2}\right)^2 (\kappa R)^4 \sin^2\theta,    \label{radiatedIntensityCircleEq}
\end{equation}
which is proportional to $\sin^2\theta$. A plot of $U \sim \sin^2\theta$ is shown in Fig.~\ref{circularBladeFig}. We can also calculate the complex power $\mathcal{P}$ given by
\[
\mathcal{P} = \oint_{\mathcal{S}} \boldsymbol{S}_o(\boldsymbol{r}) \cdot d^2\boldsymbol{r} = \int_0^{2\pi} \int_0^\pi \boldsymbol{S}_o(r,\theta) \cdot (r^2 \sin\theta d\theta d\phi \hat{r}),
\]
which is the flux of the complex Poynting vector $\boldsymbol{S}_o(r,\theta)$, given by Eqs.~(\ref{SorCircleEq})-(\ref{SophiCircleEq}), through a sphere $\mathcal{S}$ of radius $r > R$. Only the radial component of $\boldsymbol{S}_o(r,\theta)$ contributes, and the {\color{black}resulting} complex power becomes:
\[
\mathcal{P} = \frac{V_o^2}{\eta_o} (\kappa R)^4 \left[1 + \frac{1}{\mathcal{i}(\kappa r)^3}\right].
\]
Thus, the radiated power is:
\[
\mathcal{P}_{rad} = \mbox{Re}(\mathcal{P}) = \frac{V_o^2}{\eta_o} (\kappa R)^4,
\] 
which represents the energy radiated by the antenna per unit area and time. It increases with the voltage difference $V_o$ of the harmonic source and the radius $R$ of the inner sheet. Finally, we can compute the directivity:
\[
D_o := 4\pi \frac{U_{max}}{\mathcal{P}_{rad}},
\]
where $U_{max}$ is the maximum of the radiation intensity. This quantity occurs at $\theta = \pi/2$ on the outer sheet:
\[
U_{max} = U(\pi/2) = \frac{1}{2\eta_o} \left(\frac{V_o}{2}\right)^2 (\kappa R)^4 \hspace{0.5cm} \Rightarrow \hspace{0.5cm}  D_o = 4\pi \left[\frac{1}{2\eta_o} \left(\frac{V_o}{2}\right)^2 (\kappa R)^4\right] \left[\frac{\eta_o}{V_o^2} \frac{1}{(\kappa R)^4}\right] = \frac{3}{2}.
\]
{\color{black}Thus,} the directivity of the planar blade circular antenna is $D_o = 3/2${\color{black}, similar to a} wire circular loop antenna. {\color{black}This} coincidence strictly occurs in the gapless limit when the outer and inner sheets of the blade antenna are infinitely close {\color{black}to} each other.  

\begin{figure}[H]
\centering
\includegraphics[width=0.32\textwidth]{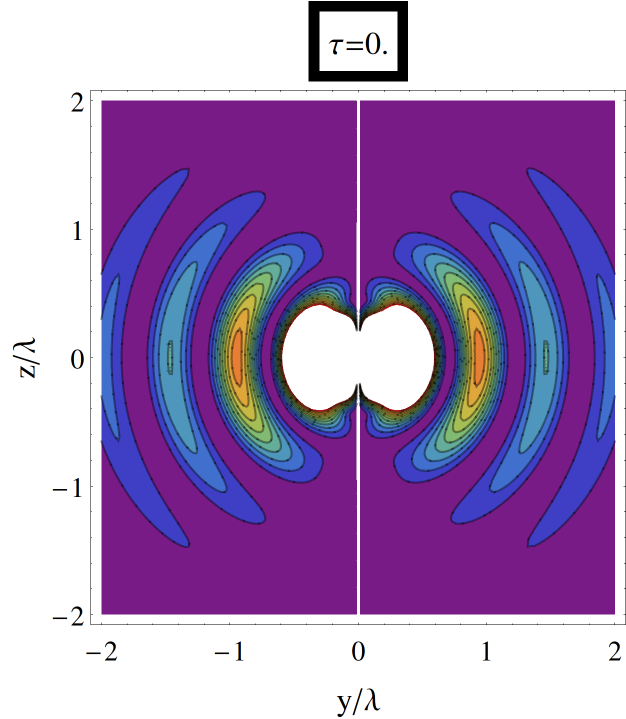}
\includegraphics[width=0.32\textwidth]{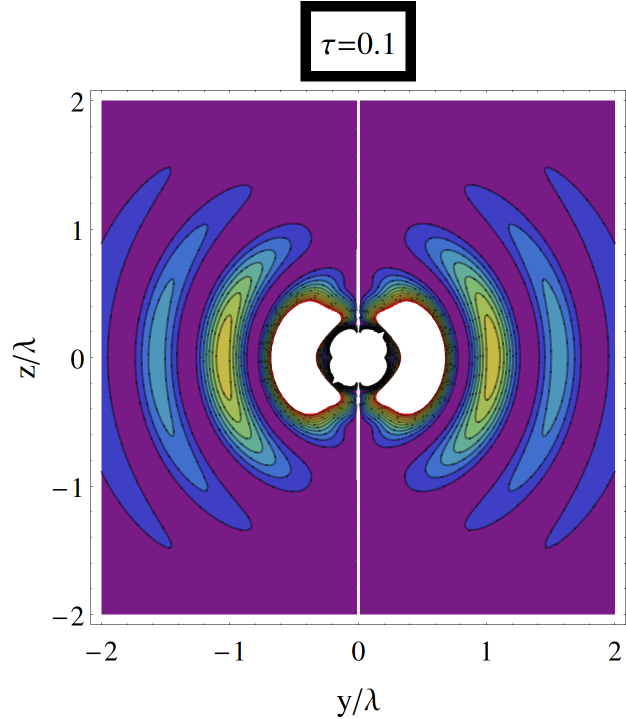}
\includegraphics[width=0.32\textwidth]{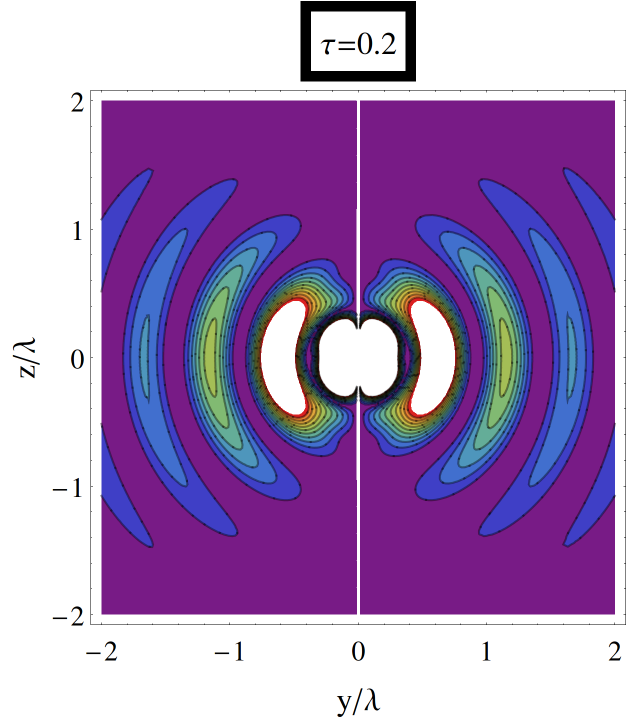}\\
\includegraphics[width=0.32\textwidth]{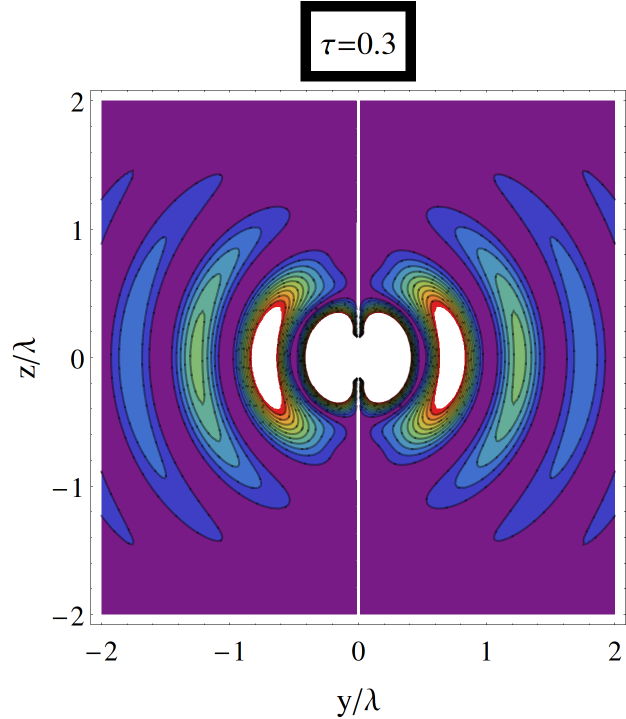}
\includegraphics[width=0.32\textwidth]{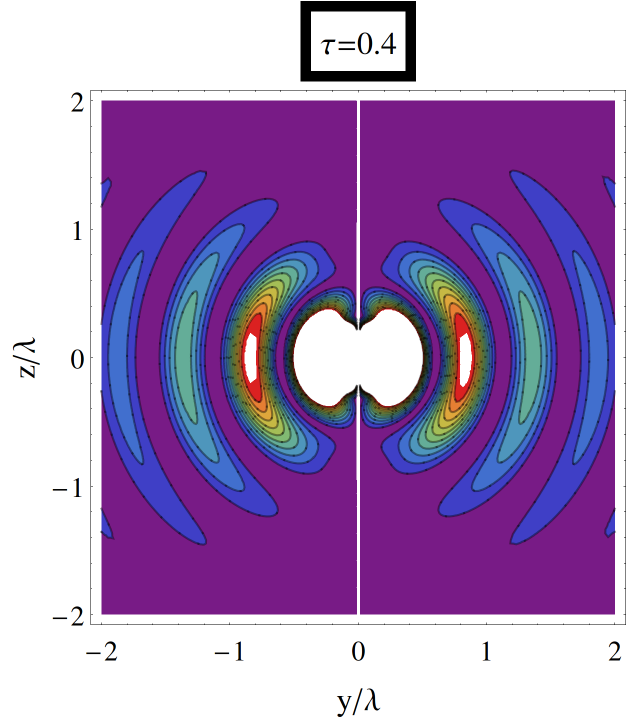}
\includegraphics[width=0.32\textwidth]{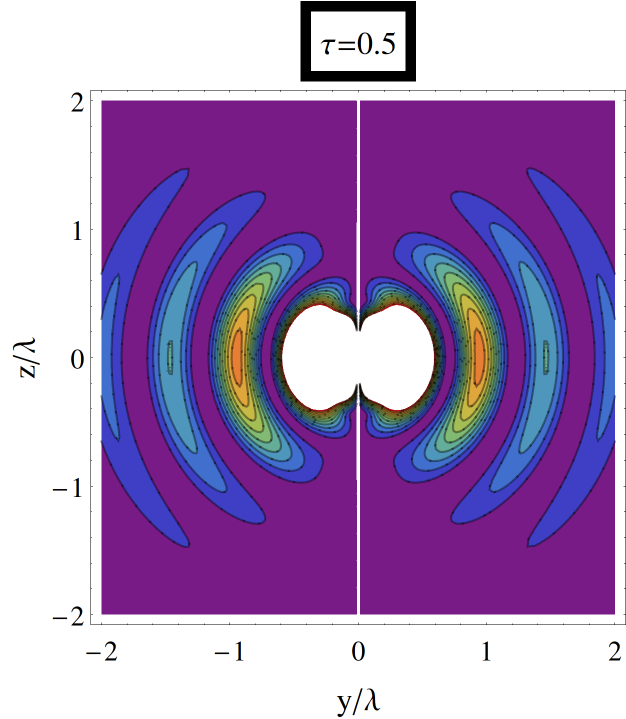}
\caption[Instantaneous Power Flow Patterns]{Magnitude of the Poynting vector $|\boldsymbol{\mathrm{S}}(\boldsymbol{r},t)|$ illustrating the instantaneous power flow patterns at different scaled times $\tau = t/T$, where $T = 2\pi/\omega$ is the period of the wave. The circular dipole blade antenna is situated on the $xy$-plane, while the visualization is on the $yz$-plane. The spatial dimensions are scaled by the wavelength $\lambda = \frac{2\pi c}{\omega}$.}
\label{SIntensityFig}
\end{figure}

The instantaneous magnetic field $\boldsymbol{\mathrm{B}}(r,t) = \mbox{Re}[\boldsymbol{B}_o(r) e^{\boldsymbol{i} \omega t}] = [\boldsymbol{B}_o(r) e^{\boldsymbol{i} \omega t} + \boldsymbol{B}_o(r)^* e^{-\boldsymbol{i} \omega t}]/2$ is given by:
\begin{equation}
\boldsymbol{\mathrm{B}}(\boldsymbol{r},t) = \mbox{sgn}(z) \frac{R^2 V_o \kappa}{2 c r^2} \sin(\theta) \left[-\kappa r \cos\left(\kappa r - \omega t\right) + \sin\left(\kappa r - \omega t\right)\right] \hat{\phi}.
\label{BInstantaneousEq}
\end{equation}

Similarly, the instantaneous electric field $\boldsymbol{\mathrm{E}}(r,t) = \mbox{Re}[\boldsymbol{E}_o(r) e^{\boldsymbol{i} \omega t}]$ is expressed as:
\begin{align}
\boldsymbol{\mathrm{E}}(\boldsymbol{r},t) &= \mbox{sgn}(z) \frac{R^2 V_o}{r^3} \cos(\theta) \left[\cos\left(\kappa r - \omega t\right) + \kappa r \sin\left(\kappa r - \omega t\right)\right] \hat{r} \nonumber \\ 
&+\mbox{sgn}(z) \frac{R^2 V_o}{2 r^3} \sin(\theta) \left[(1 - \kappa^2 r^2) \cos\left(\kappa r - \omega t\right) + \kappa r \sin\left(\kappa r - \omega t\right)\right] \hat{\theta}.
\label{EInstantaneousEq}
\end{align}

The instantaneous Poynting vector $\boldsymbol{\mathrm{S}}(\boldsymbol{r},t) = \mbox{Re}[\boldsymbol{E}_o e^{\boldsymbol{i} \omega t}] \times \mbox{Re}[\boldsymbol{B}_o e^{\boldsymbol{i} \omega t}] / \mu_o$ can be computed as follows:
\[
\boldsymbol{\mathrm{S}}(\boldsymbol{r},t) = \frac{\mathrm{B}(r,\theta,t)}{\mu_o} \left[ \mathrm{E}_\theta(r,\theta,t) \hat{r} - \mathrm{E}_r(r,\theta,t) \hat{\theta} \right],
\]
by employing Eqs.~(\ref{BInstantaneousEq}) and (\ref{EInstantaneousEq}). Plots of the magnitude of the Poynting vector $|\boldsymbol{\mathrm{S}}|$ at different times are shown in Fig.~\ref{SIntensityFig}. {\color{black}These plots illustrate} the instantaneous power flow on the $yz$-plane{\color{black}, as} this quantity exhibits axial symmetry. It is noteworthy that the radiation pattern of the circular planar dipole blade at low frequencies on the $xy$-plane (see Fig.~\ref{circularBladeFig}) bears resemblance to the dipole radiation of a vertical dipole situated along the $z$-axis. However, it should be emphasized that the $\boldsymbol{\mathrm{E}}(r,t)$ and $\boldsymbol{\mathrm{B}}(r,t)$ fields differ in both cases.

\subsubsection{Analogies between the circular blade and the wire loop antenna in the far region limit}

As previously discussed in Section~\ref{SectionAnalogiesRibbonAntennas}, the planar dipole blade and the ribbon antenna exhibit mathematical analogies within the half-real plane $\mathcal{D}$. In the gapless limit, where the layers are infinitely close to each other, the analogous systems correspond to an infinitely thin ribbon, effectively forming a wire-loop antenna. Specifically, the magnetic field of the planar dipole-blade antenna is mathematically analogous to the electric field of the wire-loop antenna. For instance, at low frequencies and in the far-field region, the electric field of the circular wire loop antenna $\boldsymbol{E}_o^{wire}(\boldsymbol{r})$ is given by (see \cite[p. 241]{balanis2016antenna}):
\begin{equation}
\boldsymbol{E}_o^{wire}(\boldsymbol{r}) = \eta \frac{(\kappa R)^2 I_o}{4} \left(1 + \frac{1}{\boldsymbol{i}\kappa r}\right) \sin\theta \frac{\exp(-\boldsymbol{i}\kappa r)}{r} \hat{\phi}(\boldsymbol{r}),
\label{ECircularWireLoopEq}
\end{equation}
where $I_o$ represents the spatial part of the electric current of the loop, assumed to be uniform. This formula for the wire circular loop is mathematically analogous to Eq.~(\ref{BFieldCircleFarEq}) in the half-plane ($z > 0$). In this analogy, the roles of electric and magnetic fields are interchanged. The key mathematical distinction between both expressions, Eq.~(\ref{ECircularWireLoopEq}) and Eq.~(\ref{BFieldCircleFarEq}), is the presence of the $\mbox{sgn}(z)$ function, which dictates the circulation of the electric field of the blade antenna: counterclockwise around the $z$-axis if $z>0$ and clockwise if $z<0$ (see Fig.~\ref{fields3DFig}-right).

Similarly, the magnetic field of the wire-loop antenna, as given by (see \cite[p. 240]{balanis2016antenna}):
\begin{align}
    B_{or}^{wire}(r,\theta) &= \frac{\kappa R^2 \mu_o I_o}{2 r^2}\left(\boldsymbol{i} + \frac{1}{\kappa r} \right)\cos\theta \exp(-\boldsymbol{i}\kappa r), \label{BorWireLoopCircleEq}\\
    B_{o\theta}^{wire}(r,\theta) &= \frac{(\kappa R)^2\mu_o I_o}{4r}\left[\frac{1}{(\kappa r)^2} + \boldsymbol{i}\frac{1}{\kappa r} - 1\right]\sin\theta \exp(-\boldsymbol{i}\kappa r), \label{BothetaWireLoopCircleEq}\\
    B_{o\phi}^{wire} &= 0, \label{BophiWireLoopCircleEq}
\end{align}
is mathematically analogous to the radiated electric field of the planar dipole blade antenna, given by Eqs.~(\ref{EorCircleEq})-(\ref{EophiCircleEq}) if $z > 0$. Once again, the roles of magnetic and electric fields are exchanged. The electric field is depicted in Fig.~\ref{fields3DFig}-left and does not generate closed loops in space, as there are charges on the sheets. Conversely, the magnetic field of its analogous counterpart, given by Eqs.~(\ref{BorWireLoopCircleEq})-(\ref{BophiWireLoopCircleEq}), generates closed loops. This occurs because the electric field of the dipole-blade antenna is not a fully solenoidal vector field due to the surface charge density and currents on the sheets, which are responsible for the mirror symmetry of the radiated fields.

Regarding the Poynting vector of the planar dipole-blade antenna, it can be concluded that $\boldsymbol{S}^{wire}$ and $\boldsymbol{S}$ differ only by a multiplicative constant, as $\boldsymbol{S}$ depends on $\mbox{sgn}(z)^2 = 1$. Therefore, the radiation patterns of the dipole blade and the circular wire-loop antenna are geometrically equivalent, and the directivity $D_o$ of both systems is exactly the same.

\subsection{Surface charge density at low frequency in the gapless limit}
In Subsection~\ref{LowFreqFarLimitRegionSeclbl}, the radiated field in the far-limit region at low frequency of the PDBA was calculated{\color{black}, confirming its consistency} with dipolar radiation. This behavior is expected, as the finite ribbon antenna (the counterpart of the PDBA) should behave like a dipole in the far-limit region. {\color{black}However, near} the PDBA's gap, the approximations employed in Subsection~\ref{LowFreqFarLimitRegionSeclbl} are no {\color{black}longer} valid, and the PDBA {\color{black}ceases to behave as} a dipole. {\color{black}In this region,} the surface charge {\color{black}density} $\sigma(\boldsymbol{r},t)$ tends to concentrate {\color{black}near the} gap boundaries $\partial \mathcal{A}_{in}$ and $\partial \mathcal{A}_{out}$. The surface charge density is proportional to the electric field evaluated on the layers. {\color{black}Thus,} the spatial part of the {\color{black}surface charge density} $\sigma_o(\boldsymbol{r})$ {\color{black}can be computed} from the electric vector potential as follows:
\begin{equation*}
\sigma_o(\kappa,u) = \frac{2 \epsilon_o}{u} \lim_{\theta \to \frac{\pi}{2}} \frac{\partial}{\partial r} \left[r (\Theta_o)_\phi(\kappa,r,\theta)\right],
\end{equation*}
where $u=\sqrt{x^2 + y^2}$, assuming low-frequency radiation. In general, for the gapless limit, one must solve Eq.~(\ref{ThetaWireLikeEq}) to evaluate the electric vector potential. The {\color{black}challenge is determining} the potential $\mathcal{V}_{12}(\boldsymbol{r})$ {\color{black}along} the mid-trajectory $c_{mid}$, as it can {\color{black}vary based} on the loop parameter, the value of $\kappa$, and the position {\color{black}of the voltage feed} on $c_{mid}$.

\begin{figure}[H]
\centering
\includegraphics[width=0.35\textwidth]{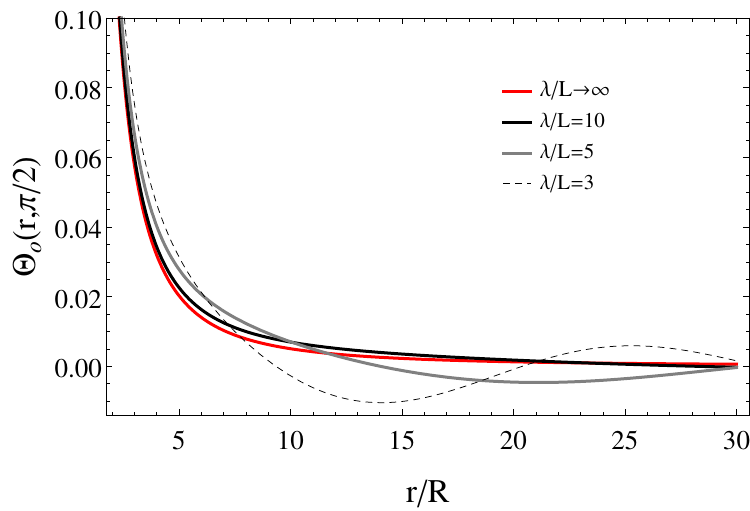}
\includegraphics[width=0.29\textwidth]{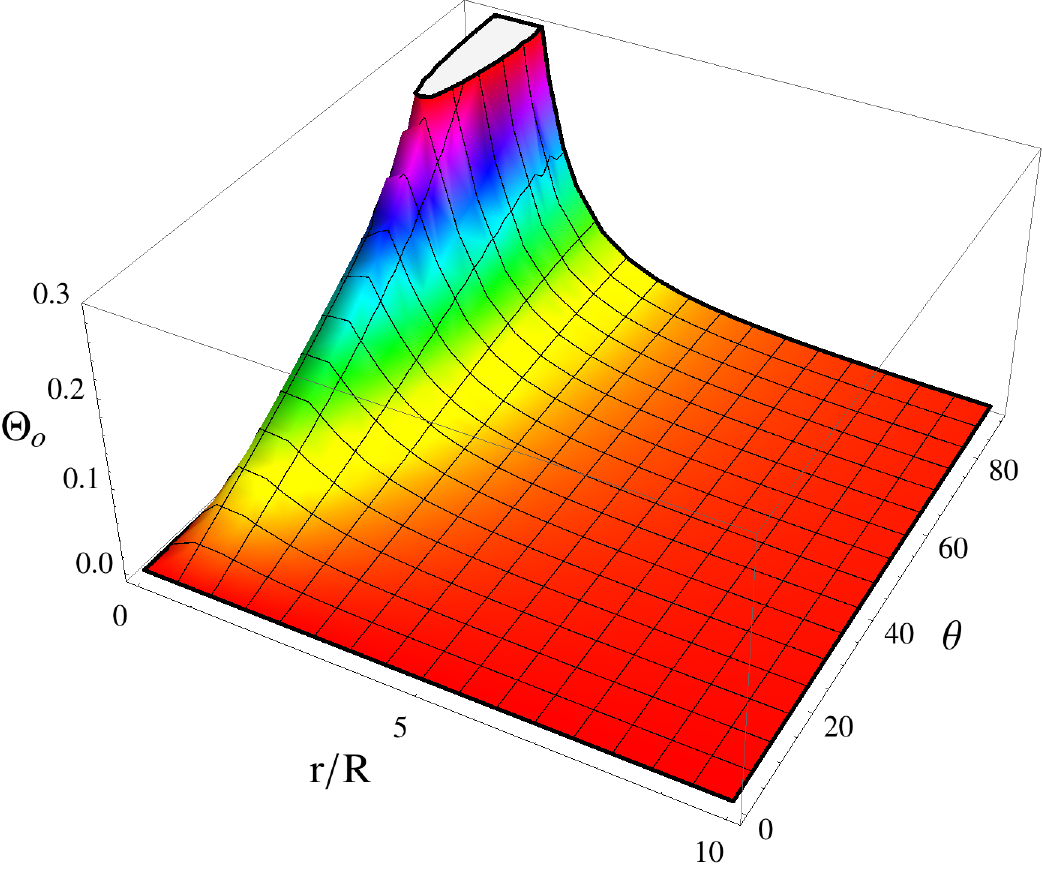}
\includegraphics[width=0.29\textwidth]{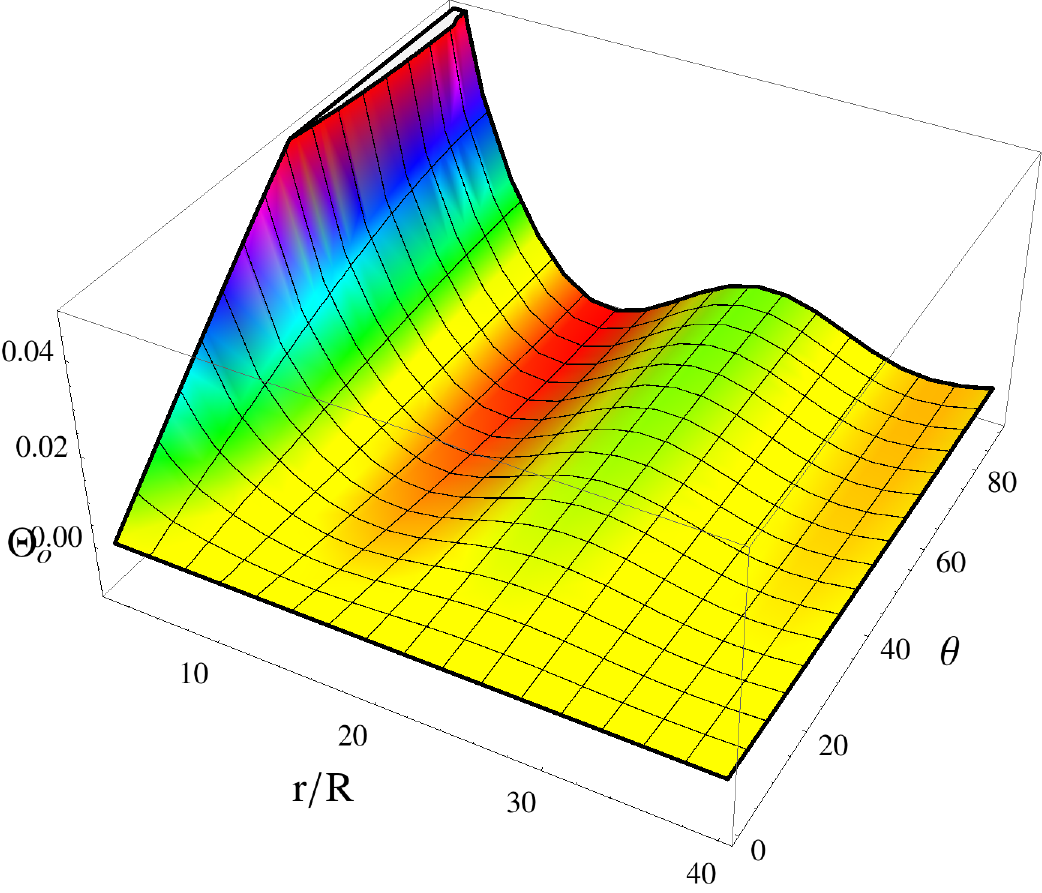}
    \caption[Electric Vector Potential]{Electric vector potential: (Left) $\Theta$ evaluated on the outer layer. Electric vector potential in space for (center) $\lambda/L\rightarrow\infty$ and (right) $\lambda/L=3$.}
\label{Theta3DFig}
\end{figure}

\begin{figure}[H]
\centering
\includegraphics[width=0.35\textwidth]{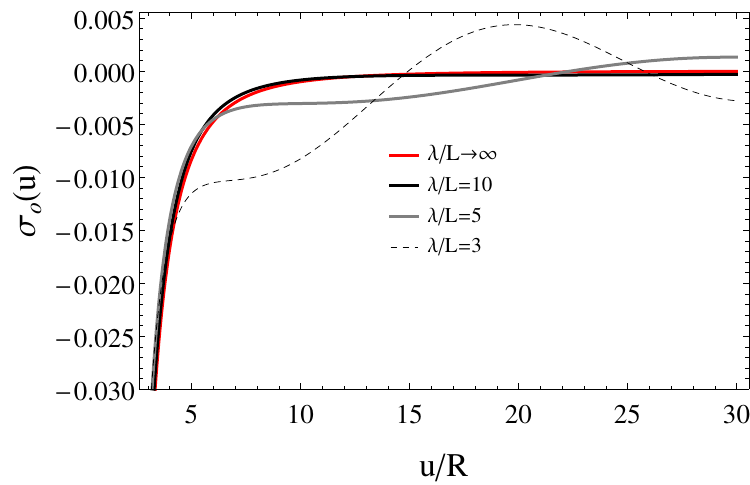}
\includegraphics[width=0.29\textwidth]{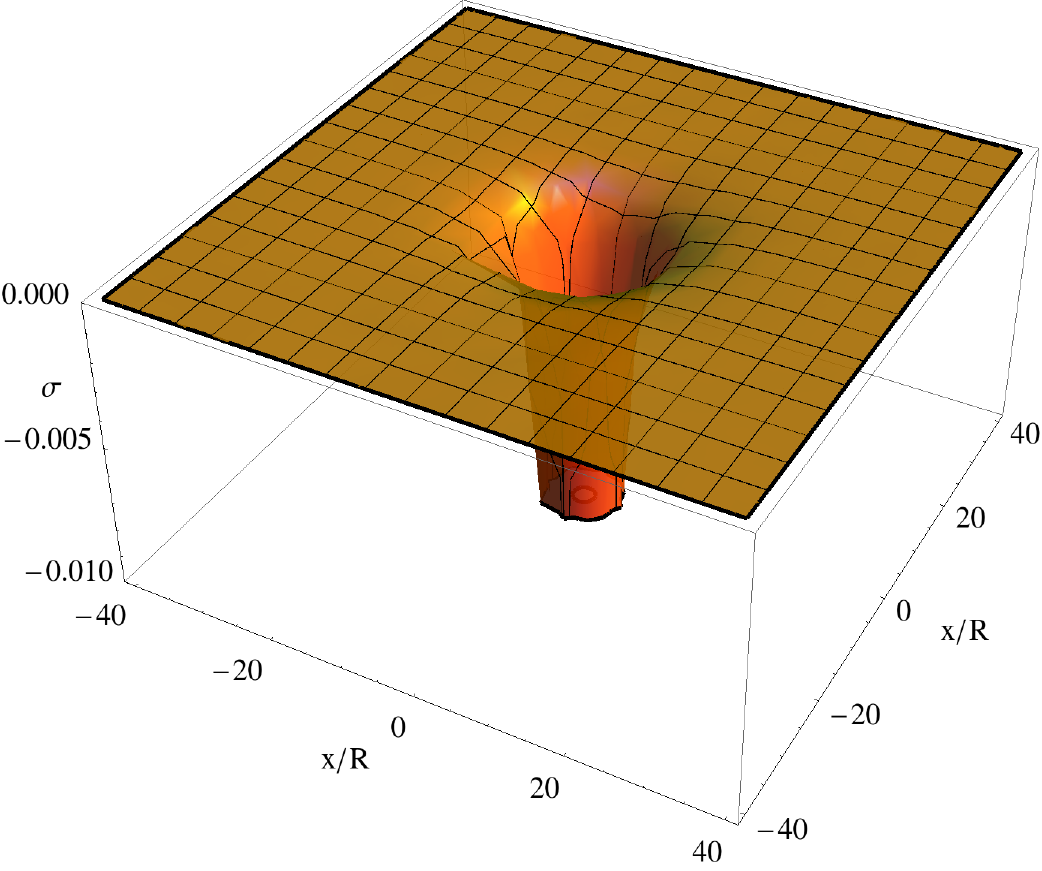}
\includegraphics[width=0.29\textwidth]{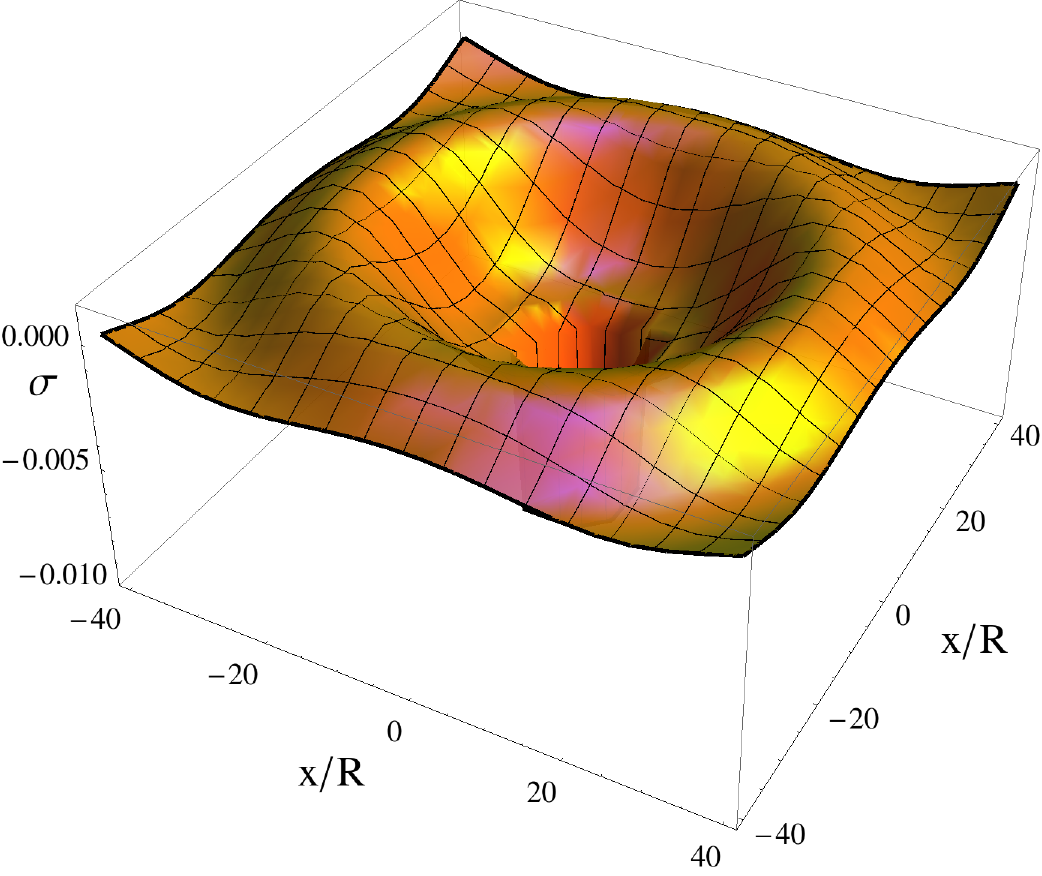}
    \caption[Surface charge density]{Surface charge density of the outer layer: (Left) $\sigma$ evaluated on the outer layer; surface charge density for (center) $\lambda/L \rightarrow \infty$ and (right) $\lambda/L = 3$.}
\label{Sigma3DFig}
\end{figure}

In the limit $\kappa = \omega/c \to 0$, we know that $\mathcal{V}_{12}(\boldsymbol{r}) = V_o$ is constant, with $V_o$ {\color{black}being} the potential difference between layers{\color{black}, and the} location of the voltage source {\color{black}is negligible}. At low frequency, $\mathcal{V}_{12}(\boldsymbol{r})$ can be assumed uniform if the emitted wavelength $\lambda$ is {\color{black}much larger than} the typical length $L$ of the PDBA's gap. It is well-known that current profiles on wire circular loop antennas are approximately uniform {\color{black}when} the perimeter of the antenna is less than about $\lambda/5$ \cite{balanis2016antenna}. In this subsection, we assume that $\lambda$ is sufficiently {\color{black}large} to consider $\mathcal{V}_{12}(\boldsymbol{r})$ uniform along $c_{mid}$. {\color{black}Considering this,} the electric vector potential can be evaluated from
\begin{equation}
(\Theta_o)_\phi^{\mbox{gapless}}(\kappa,\boldsymbol{r}) = \frac{V_o R}{2\pi} \int_{0}^{2\pi} \frac{d\phi'}{\mathcal{r}(r,\theta,\phi-\phi')} \cos(\phi-\phi') \exp(-\boldsymbol{i} \kappa \mathcal{r}(r,\theta,\phi-\phi')),
\label{ThetaSymmmetricIntegralEq}
\end{equation}
{\color{black}where} we can choose $\phi = 0${\color{black}, given the} expected azimuthal symmetry. The exact solution in the electrostatic limit {\color{black}corresponds to} the gapless SE \cite{salazar2020gaped} of radius $R$:
\[
(\Theta_o)_\phi(\boldsymbol{r},R)^{\mbox{GSE}} = \lim_{\kappa \to 0} (\Theta_o)_\phi(\kappa,\boldsymbol{r}) = \frac{V_o}{2\pi} \frac{4R}{\sqrt{R^2 + r^2 + 2Rr\sin\theta}} \left[\frac{(2-\gamma^2)K(\gamma^2) - 2\underbar{E}(\gamma^2)}{\gamma^2}\right],
\]
where $\gamma^2(r,\theta) := \frac{4Rr\sin\theta}{R^2 + r^2 + 2Rr\sin\theta}$, and $K$ and $\underbar{E}$ are the complete elliptic integrals of the first and second kind, respectively. The static surface charge density of the layers is:
\[
\lim_{\kappa \to 0} \sigma_o(\kappa,u) = \frac{2\epsilon_o V_o}{\pi} \left[ \frac{1}{R-u} \underbar{E}\left(\frac{4Ru}{(R+u)^2}\right) + \frac{1}{R+u} K\left(\frac{4Ru}{(R+u)^2}\right)\right].
\]
These static exact solutions are shown in Figs.~\ref{Theta3DFig} and \ref{Sigma3DFig} for $\kappa = \frac{2\pi}{\lambda} \to 0$, or equivalently $\lambda/L \to \infty$, where $\lambda$ is the wavelength and $L = 2\pi R$. As the radiated wavelength decreases, the profile of the electric vector potential changes, as shown in Fig.~\ref{Theta3DFig}, similarly for the surface charge density (see Fig.~\ref{Sigma3DFig}). For larger values of $\lambda$ (e.g., $\lambda/L > 10$), the total charge on the layers is practically interchanged in half a period $T/2$, with $T = 2\pi/\omega$, leading to charges of the same sign on each layer. {\color{black}Conversely, when} $\lambda \approx 5L$, the outer layer {\color{black}exhibits} a mixture of positive and negative charges distributed in concentric annular regions. {\color{black}If} the wavelength approaches the typical length of the gap ($\lambda \simeq L$), the surface charge density profiles become more complex{\color{black}, as the system is} no longer axially symmetric.

\subsection{Annular Gapped Dipole Radiator at Low Frequency}
In this section, we limit the study to the case $\kappa\mathcal{r}(r,\theta,\phi') \ll 1$, {\color{black}focusing on} the inter-layer region at low frequency. {\color{black}Expanding the exponential term} $\exp(-\boldsymbol{i}\kappa \mathcal{r})$ in Eq.~(\ref{ThetaSymmmetricIntegralEq}){\color{black}, we can express the electric vector potential as:}
\begin{equation}
(\Theta_o)_\phi^{\mbox{gapless}}(\kappa,\boldsymbol{r},R) = (\Theta_o)_\phi(\boldsymbol{r},R)^{\mbox{GSE}}(r,\theta,R) + \frac{V_o R}{2\pi} \sum_{n=1}^\infty \frac{(-\boldsymbol{i}\kappa)^n}{n!} I_n(r,\theta,R), 
\label{gaplessThetaExpansionEq}
\end{equation}
with
\[
I_n(r,\theta,R) = (r^2+R^2)^{n/2} \int_0^{2\pi} \cos(\phi') \left(1-\xi\cos(\phi')\right)^{n/2} d\phi', 
\]
and $\xi := \frac{2rR\sin\theta}{r^2+R^2}$ in the limit $\Delta R \to 0$.

The real part of {\color{black}the} magnetic field $\boldsymbol{B}(\boldsymbol{r},t) = \mbox{Re}[\boldsymbol{i}\frac{\kappa}{c} \boldsymbol{\Theta}_o \exp{\left(\boldsymbol{i}\omega t\right)}]$ at first order in $\kappa$ is:
\[
\boldsymbol{B}^{\mbox{gapless}}(\boldsymbol{r},t) = -\frac{\kappa}{c} (\Theta_o)_\phi(\boldsymbol{r})^{\mbox{GSE}} \sin(\omega t) \hat{\phi} + O(\kappa^2),
\]
leading to:
\[
\boldsymbol{B}^{\mbox{gapless}}(\boldsymbol{r},t) = -\mbox{sgn}(z)\frac{\mathcal{B}_o(\kappa)}{2\pi}\frac{4R \sin(\omega t)}{\sqrt{R^2+r^2+2Rr\sin\theta}}\left[\frac{(2-\gamma^2)K(\gamma^2)-2\underbar{E}(\gamma^2)}{\gamma^2}\right]\hat{\phi} + O(\kappa^3),
\]
with $\mathcal{B}_o(\kappa) = \frac{\kappa V_o}{c}$ as the scaling factor with units of magnetic field.

Similarly, the electric field is $\boldsymbol{E}(\boldsymbol{r},t) = \nabla\times\Theta^{GSE} \cos(\omega t) + O(\kappa^2)$. {\color{black}This results in the following components of the electric field:}
\begin{align}
    E_r^{\mbox{gapless}}(\boldsymbol{r},t) &= \frac{R^2 V_o}{2\pi} \frac{4\cos\theta\cos(\omega t)}{\mathscr{r}_{-}(r,\theta)^2 \mathscr{r}_{+}(r,\theta)} \underbar{E}\left( \frac{4rR\sin\theta}{ \mathscr{r}_{+}(r,\theta)^2 }\right) + O(\kappa^2), \\
    E_\theta^{\mbox{gapless}}(\boldsymbol{r},t) &= \frac{V_o}{\pi} \frac{\csc\theta \cos(\omega t)}{\mathscr{r}_{-}^2 \mathscr{r}_{+}} \left[ (r^2+R^2\cos(2\theta))\underbar{E}\left( \frac{4rR\sin\theta}{ \mathscr{r}_{+}^2 }\right) - \mathscr{r}_{-}^2 K\left( \frac{4rR\sin\theta}{ \mathscr{r}_{+}^2 }\right)\right] + O(\kappa^2), \\
    E_\phi^{\mbox{gapless}}(\boldsymbol{r},t) &= 0,
\end{align}
where $\mathscr{r}_{\pm}(r,\theta) = \sqrt{r^2 + R^2 \pm 2rR\sin\theta}$.

\begin{figure}[H]
\centering
\includegraphics[width=0.35\textwidth]{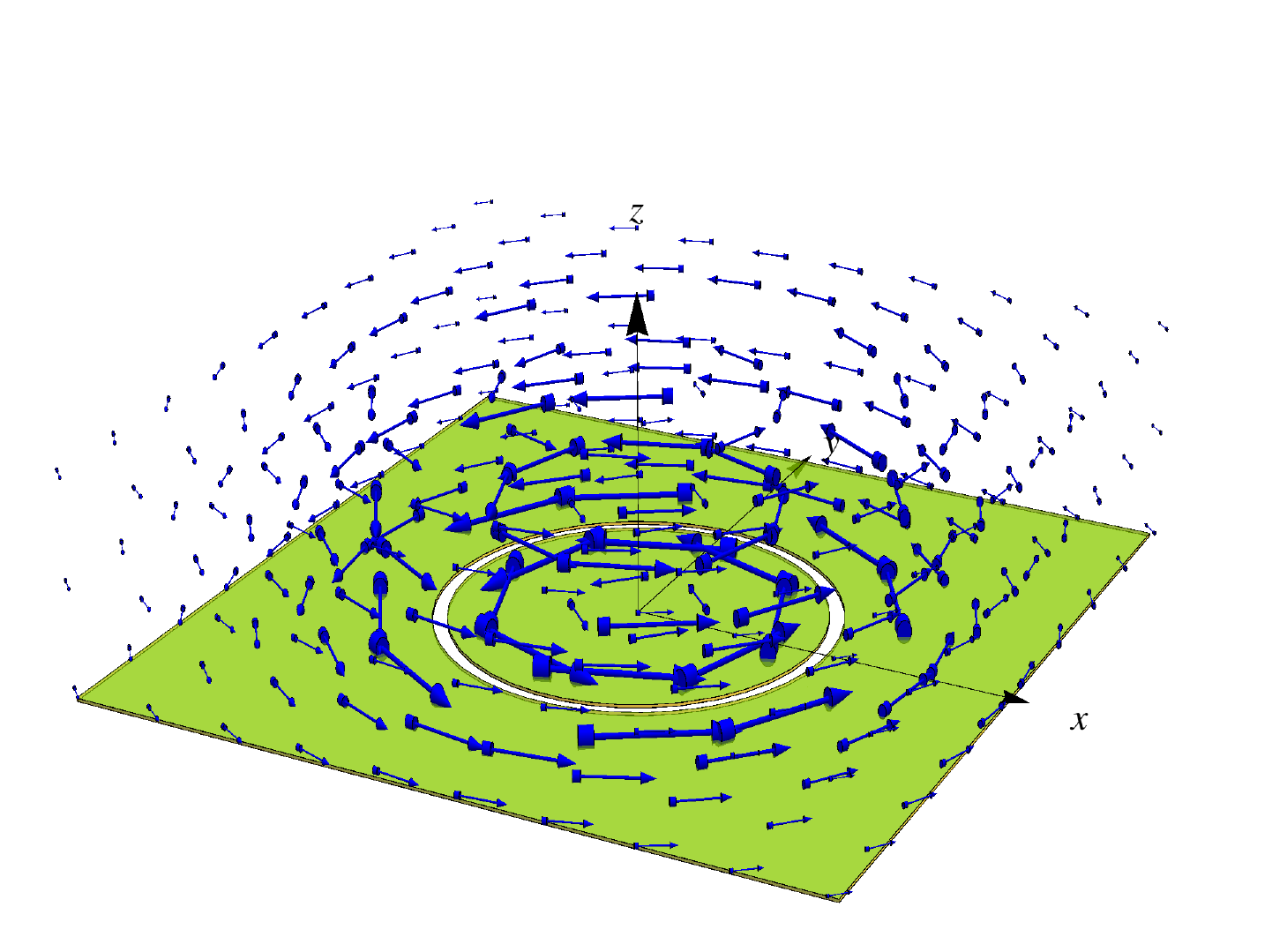}
\includegraphics[width=0.29\textwidth]{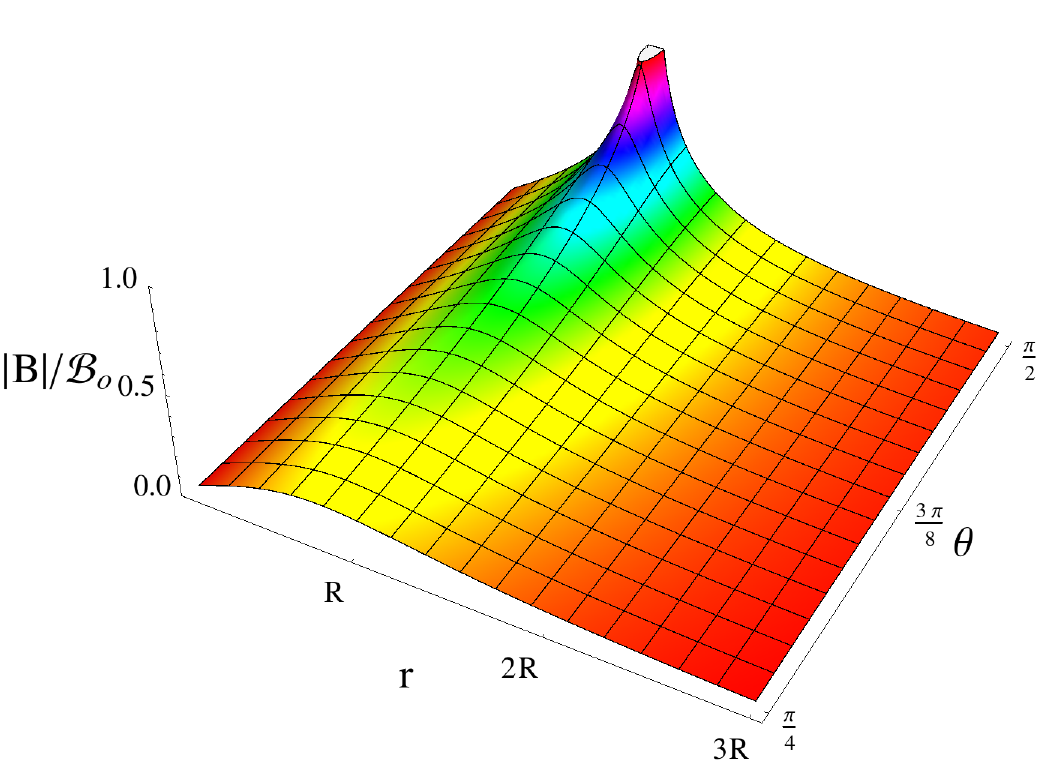}
\includegraphics[width=0.29\textwidth]{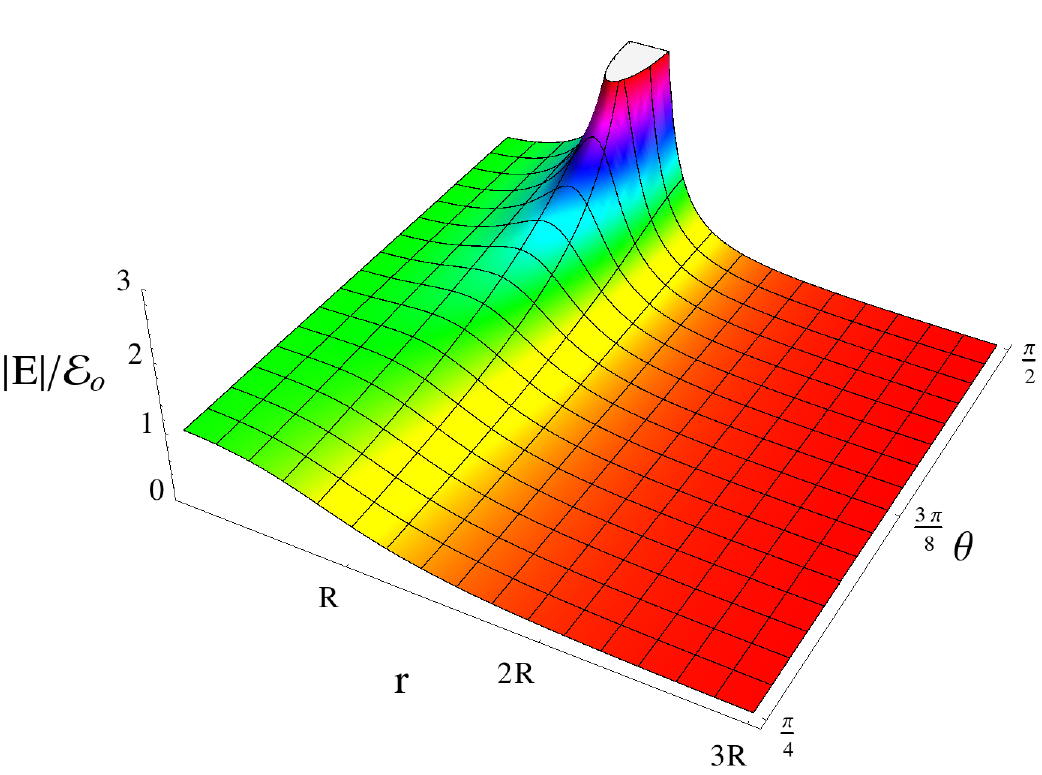}
    \caption[Fields near the gap]{Vector plot of the magnetic field at $t=5T/8$. Magnitude of the (center) magnetic field and (right) electric field. The scale factors are $\mathcal{B}_o = \frac{\kappa V_o}{c}$ and $\mathcal{E}_o = \frac{V_o}{R}$.}
\label{FieldsNear3DFig}
\end{figure}

{\color{black}Fig.~\ref{FieldsNear3DFig} shows a} plot of the fields near the inter-layer region in the limit $\Delta R \to 0$. {\color{black}The} magnetic and electric fields are more intense as the point of evaluation {\color{black}approaches} the inter-layer region $\mathcal{r}(r,\theta,\phi') \to 0$ for $z > 0$.

{\color{black}Next, we} consider the gapped case. The electric vector potential can be evaluated from Eq.~(\ref{electricVectoPotentialFinalEq}) with $\mathcal{G}$ {\color{black}as an} annular region of thickness $\Delta R$:
\[
\boldsymbol{\Theta}(\boldsymbol{r},t) = \frac{1}{2\pi} \int_{\mathcal{G}\in\mathbb{R}^2} \frac{1}{|\boldsymbol{r}-\boldsymbol{r}'|} \boldsymbol{\mathcal{W}}(u',\phi') \exp\left(-\boldsymbol{i} \frac{\omega |\boldsymbol{r}-\boldsymbol{r}'|}{c}\right) u'du'd\phi'.
\]
If $\kappa = 2\pi/\lambda$ is sufficiently small (say $\lambda > 10 L$), azimuthal symmetry {\color{black}can be assumed, implying} that $|\boldsymbol{\mathcal{W}}(\boldsymbol{r})| = \mathcal{W}(u)$. The weight vector becomes:
\[
\boldsymbol{\mathcal{W}}(\boldsymbol{r}') = \mathcal{W}_\phi(u')\left[\sin\theta\sin(\phi-\phi')\hat{r}(\boldsymbol{r}) + \cos\theta\sin(\phi-\phi')\hat{\theta}(\boldsymbol{r}) + \cos(\phi-\phi')\hat{\phi}(\boldsymbol{r})\right],
\]
with $\mathcal{r}(\boldsymbol{r},u',\phi') = |\boldsymbol{r}-\boldsymbol{r}'| = \sqrt{r^2+u'^2-2ru'\sin\theta\cos(\phi-\phi')}$. {\color{black}This reduces} the electric vector potential to:
\[
\boldsymbol{\Theta}(\boldsymbol{r},t) = \frac{1}{2\pi} \int_{R-\frac{\Delta R}{2}}^{R+\frac{\Delta R}{2}} \mathcal{W}_\phi(u') u'du' \int_0^{2\pi} d\phi' \frac{1}{\mathcal{r}(\boldsymbol{r},u',\phi')} \exp\left(-\boldsymbol{i} \kappa \mathcal{r}(\boldsymbol{r},u',\phi')\right) \cos(\phi-\phi')\hat{\phi}(\boldsymbol{r}).
\]
Rearranging the terms inside the integral:
\[
\boldsymbol{\Theta}(\boldsymbol{r},t) = \int_{R-\frac{\Delta R}{2}}^{R+\frac{\Delta R}{2}} \frac{\mathcal{W}_\phi(u')}{V_o} \left\{\frac{V_o u'}{2\pi} \int_0^{2\pi} \frac{d\phi'}{\mathcal{r}(\boldsymbol{r},u',\phi')} \exp\left(-\boldsymbol{i} \kappa \mathcal{r}(\boldsymbol{r},u',\phi')\right) \cos(\phi-\phi') \hat{\phi}(\boldsymbol{r}) \right\} du'.
\]

\begin{figure}[H]
\centering
\includegraphics[width=0.95\textwidth]{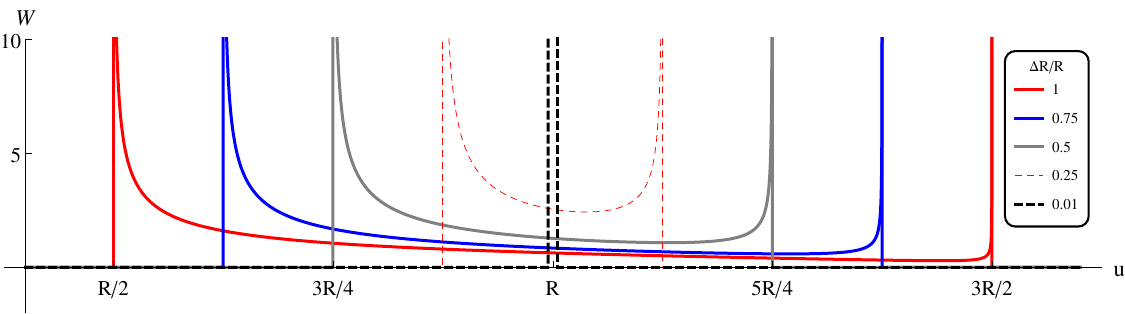}
    \caption[Weight Vector]{Vector $\mathcal{W}_\phi(u)$ as a function of the gap.}
\label{WVectorFig}
\end{figure}

Hence, one can identify the gapless contribution inside the radial integral (see Eq.~(\ref{ThetaSymmmetricIntegralEq})) to write  
\[
\boldsymbol{\Theta}(\boldsymbol{r},t)=\int_{R-\frac{\Delta R}{2}}^{R+\frac{\Delta R}{2}} \frac{\mathcal{W}_\phi(u')}{V_o}(\Theta_o)_\phi^{\mbox{gapless}}(\kappa,\boldsymbol{r},u') du'
\]
and expanding according to Eq.~(\ref{gaplessThetaExpansionEq}) gives:
\[
\boldsymbol{\Theta}(\boldsymbol{r},t)=\int_{R-\frac{\Delta R}{2}}^{R+\frac{\Delta R}{2}} \frac{\mathcal{W}_\phi(u')}{V_o}(\Theta_o)_\phi(\boldsymbol{r},R)^{\mbox{GSE}}(r,\theta,u')du' + \frac{1}{2\pi} \sum_{n=1}^\infty \frac{(-\boldsymbol{i}\kappa)^n}{n!} \int_{R-\frac{\Delta R}{2}}^{R+\frac{\Delta R}{2}}\mathcal{W}_\phi(u')I_n(r,\theta,u')u'du'.
\]

\begin{figure}[H]
\centering
\includegraphics[width=0.32\textwidth]{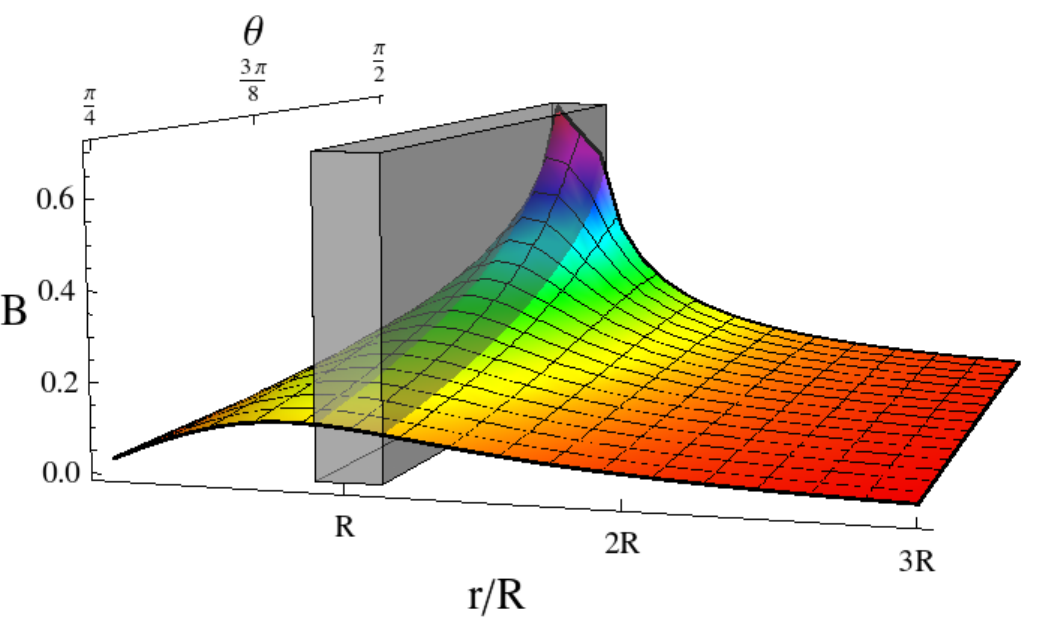}
\includegraphics[width=0.32\textwidth]{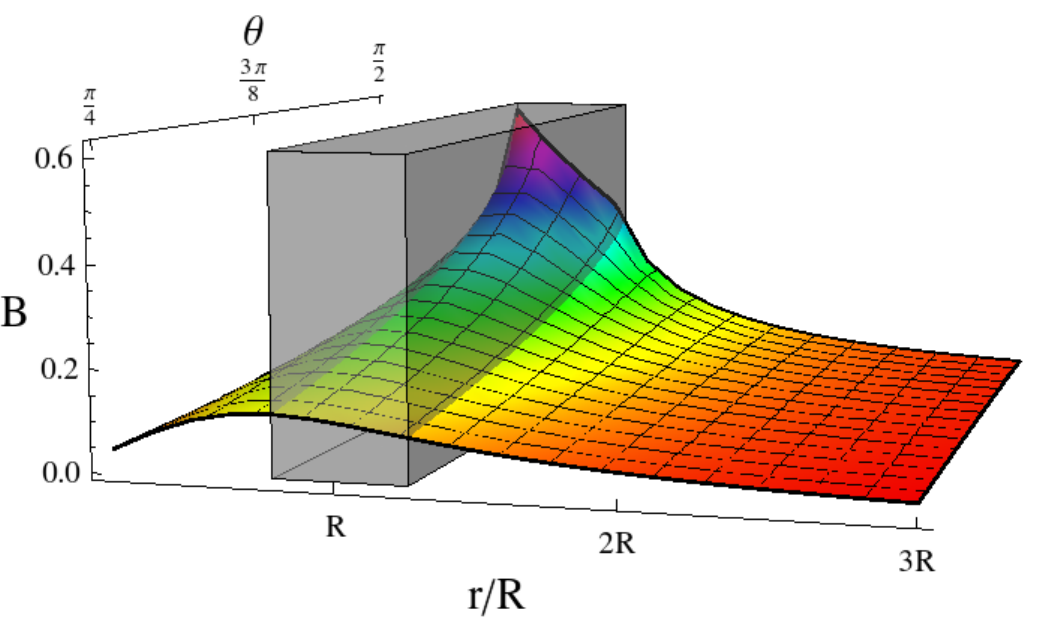}
\includegraphics[width=0.32\textwidth]{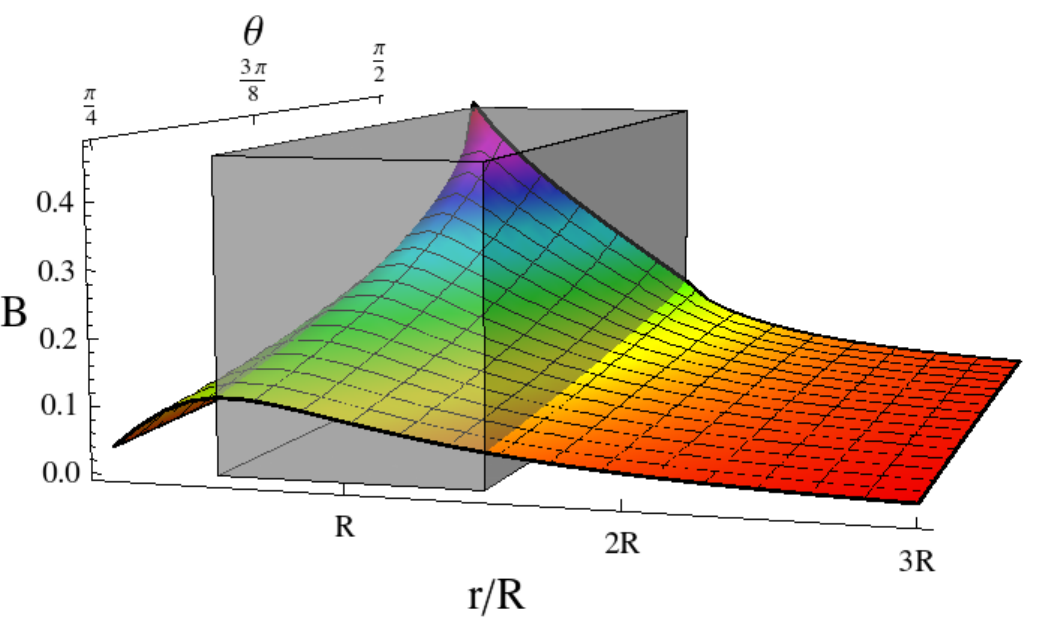}\\ 
\includegraphics[width=0.28\textwidth]{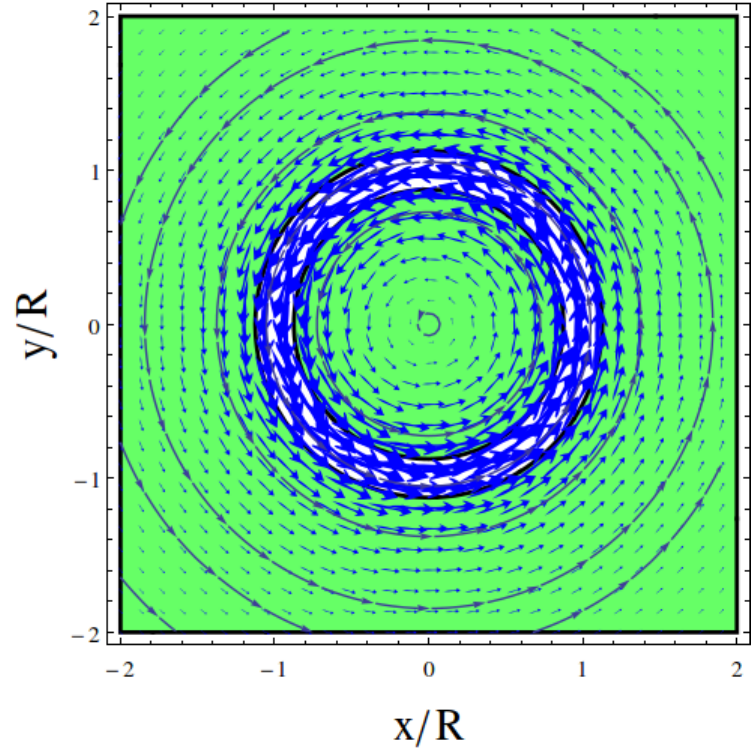}\hspace{0.5cm}
\includegraphics[width=0.28\textwidth]{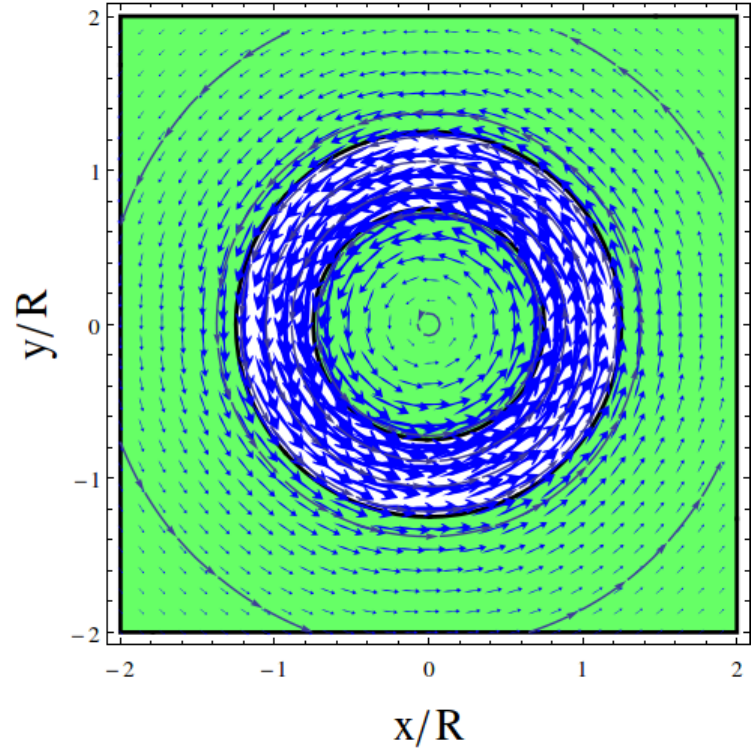}\hspace{0.5cm}
\includegraphics[width=0.28\textwidth]{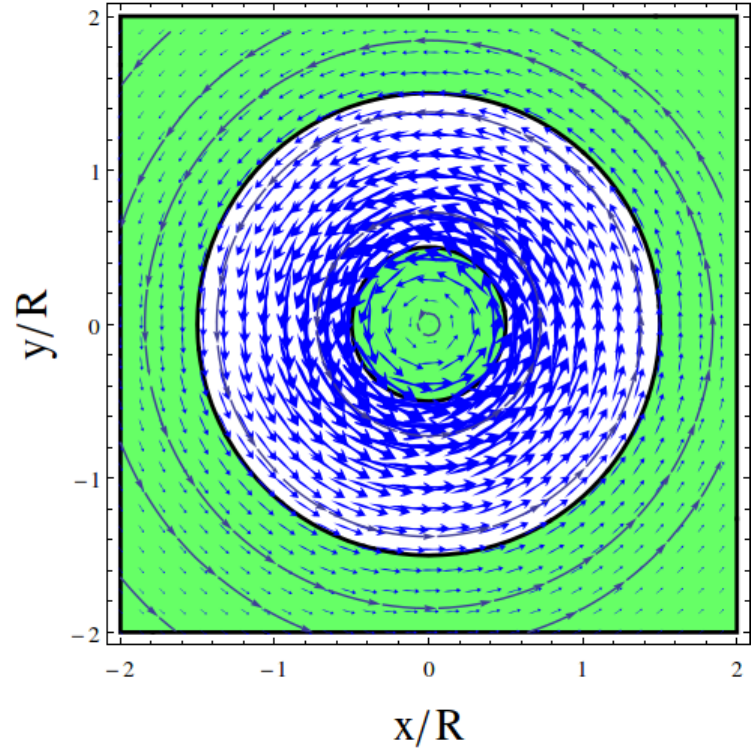}
    \caption[The system.]{Magnetic field at $z\rightarrow 0^+$. The shadowed regions represent the gap, set as (left) $\nu=\Delta R/R = 0.25$, (center) $\nu= 0.5$, and (right) $\nu=\Delta R/R = 1$.}
\label{BFieldGappedFig}
\end{figure}

The weight vector $\mathcal{W}_\phi(u')$ can be computed using the same techniques described in \cite{schmied2010electrostatics, salazar2020gaped, romero2022monte, schmied2010electrostatics} for the circular surface electrode. A plot of the weight vector as a function of the gap thickness $\Delta R$ is shown in Fig.~\ref{WVectorFig}. The spatial magnetic field in the near region at low frequency is shown in Fig.~\ref{BFieldGappedFig} in the limit $z\rightarrow 0^{+}$. 

\section{Limitations and future research}
In this study, a mathematical equivalence between radiation from a PDBA and current ribbon radiation was formally demonstrated through the dual representation of the electromagnetic field. {\color{black}This necessitated constructing} the Green's function, taking into account the boundary conditions of the system to solve the wave equation in the dual form. The Green's function presented in \newtext{this paper} is strictly correct if the outer layer $\mathcal{A}_{out}$ is semi-infinite. In practice, {\color{black}extrapolating} this technique for a finite antenna (e.g., if $\mathcal{A}_{out}$ is another annular region for the circular case) would {\color{black}require constructing a new} Green's function that correctly adapts to the boundary conditions. This is not a trivial task, and the mathematical equivalence with current ribbons {\color{black}might} no longer apply.

Another {\color{black}scenario} to consider is the case of high-frequency radiation. In such a {\color{black}situation}, the weight vector $\boldsymbol{\mathcal{W}}$ should be calculated in the region between the plates to compute the electric vector potential. This {\color{black}would likely need to} be done numerically, for example, by adapting the Method of Moments (MoM) \cite{gibson2021method, li2004method, harrington1993, rao1982, chew1995, jin2002} to determine the weight vector instead of the currents and charge on the plates. In theory, this would be more computationally economical because the gap is finite compared to the metallic plates. The {\color{black}challenge with} this strategy {\color{black}lies} in the nature of the feeding {\color{black}mechanism, as} the type of voltage source and its location typically impact {\color{black}high-frequency behavior.} This {\color{black}is} a situation to {\color{black}explore} in future work.

\section{Conclusion}
In \newtext{this paper}, {\color{black}we} studied a planar dipole blade antenna using a formalism based on the dual electromagnetic 4-potential representation, emphasizing the association of a vector potential $\boldsymbol{\Theta}(\boldsymbol{r},t)$ with the electric field $\boldsymbol{E}=\nabla\times\boldsymbol{\Theta}$ within the domain $\mathcal{D}$. This choice of representation naturally leads to the D'Alembert equation when an appropriate Gauge condition, akin to the Lorentz Gauge condition, is applied.

We demonstrated the feasibility of obtaining integral solutions for the D'Alembert equation governing the electric vector potential emitted by a specific configuration of a dipole-blade antenna by determining the Green's function of the system. This analysis {\color{black}highlighted} the mathematical equivalence between certain radiation systems involving planar surfaces in the half-real space $\mathcal{D}$ and ribbon-loop antennas deployed in the $\mathcal{G}$ region between plates. Consequently, the electric vector potential $\boldsymbol{\Theta}$ of the dipole-blade antenna becomes mathematically analogous to the magnetic vector potential $\boldsymbol{A}$ of a ribbon-loop antenna within the half-real space.

The significance of this formalism {\color{black}is evident in} its potential applications {\color{black}across} various technological domains. Our analytical results could find immediate {\color{black}application} in radar systems, antennas, wireless communication systems, satellite communications, and IoT devices. Understanding the intricate details of radiation patterns and field distributions in complex technological configurations is crucial. Our analytical tools offer an efficient and cost-effective means of characterizing and designing radiating systems, {\color{black}providing valuable support for} engineering teams developing advanced applications. {\color{black}By leveraging} the interplay between vector potentials and electric and magnetic fields, engineers and physicists can {\color{black}use this formalism} to advance their research and design efforts, ultimately driving innovation and progress in numerous fields.

\section*{Acknowledgments}
This work was partially supported by Vicerrector\'ia de investigaci\'on, Universidad ECCI. 
{\color{black}C. Bayona-Roa acknowledges the technical support provided by the Centro de Ingenier\'ia Avanzada y de Investigaci\'on y Desarrollo -- CIAID.}

\section*{Data Availability Statement}
No data is associated with this manuscript.

\bibliographystyle{ieeetr} 
\bibliography{bibliography.bib}

\begin{appendices}

\section{Lorentz invariance}
\label{Lorentzappendix}
A fundamental requirement for the application of the 4-dual potential representation presented in Eq.~(\ref{dual4PotentialEq}) is that the laws involving $\Theta^\mu$ must remain invariant under Lorentz transformations. Without this invariance, the 4-dual potential would not accurately represent electromagnetic radiation. The Lorentz transformation, in its simplest form for inertial frames, is given by

\begin{align}
x'^\mu = (ct', \boldsymbol{x}') &= \begin{pmatrix}
           \gamma(ct-\beta x) \\
           \gamma(x-\beta ct) \\
           y \\
           z
         \end{pmatrix},
\end{align}         

where $\beta=v/c$ and $\gamma=1/\sqrt{1-\beta^2}$ are the velocity and Lorentz factor, respectively. This linear position transformation can be expressed in tensor form as $x'^\mu=\Lambda^\mu_\nu x^\nu$, which generalizes transformations between contravariant vectors, including the 4-dual potential in Eq.~(\ref{dual4PotentialEq}). Specifically,
\begin{align}
\Theta'^\mu = \Lambda^{\mu}_{\nu} \Theta^\nu &= 
\begin{pmatrix}
\gamma &-\beta\gamma &0 &0\\
-\beta\gamma &\gamma &0 &0\\
0 &0 &1 &0\\
0 &0 &0 &1
\end{pmatrix}
\begin{pmatrix}
           \mu_0\Psi/c \\
           -\Theta_x/c^2 \\
           -\Theta_y/c^2 \\
           -\Theta_z/c^2
         \end{pmatrix}.
         \label{lorentzTransEq}
\end{align}         
In more explicit terms, this transformation yields: 
\[
\Psi' = \gamma\left(\Psi+\frac{\beta}{\mu_0 c}\Theta_x\right), \hspace{0.5cm}\Theta'_x = \gamma\left(\mu_0 v\Psi+\Theta_x\right),\hspace{0.5cm}\Theta'_y=\Theta_y,\hspace{0.5cm}\Theta'_z=\Theta_z.
\]
These represent the components of the 4-dual potential in the moving inertial frame. Our goal is to investigate the behavior of the d'Alembert operator under this transformation. The d'Alembert operator is given by
\[
    \Box'^2= -\frac{1}{c^2}\partial^2_{t'}-\partial_{x'}^2-\partial_{y'}^2-\partial_{z'}^2 
\]    
where each term $\partial_{t'}=\gamma(\partial_t+v\partial_x)$, $\partial_{x'}=\gamma(\partial_x+v/c^2\partial_t)$, $\partial_{y'}=\partial_y$, and $\partial_{z'}=\partial_z$. Second derivatives become $\partial_{t'}^2 = \gamma^2(\partial_t^2+v^2\partial_x^2+v\partial_x\partial_t)$ and $\partial_{x'}^2 = \gamma^2(\partial_x^2+v^2/c^2\partial_t^2+v/c\partial_x\partial_t)$. Consequently,
\begin{align*}
\Box'^2 &= \left(\frac{\gamma^2}{c^2}-\frac{\gamma^2\beta^2}{c^2}\right)\partial^2_{t}+\left(\frac{\gamma^2v^2}{c^2}-\gamma^2\right)\partial^2_{x'}+ \left(\frac{\gamma^2v}{c^2}-\frac{\gamma^2\beta}{c}\right)\partial_x\partial_t +\partial^2_{y}-\partial^2_{z} \\
  &=\frac{1}{c^2}\partial^2_{t}-\partial^2_{x}-\partial^2_{y}-\partial^2_{z} = \Box^2
\end{align*}
which demonstrates that the d'Alembert operator is invariant under the Lorentz transformation. Consequently, the 4-dimensional wave equation holds:
\[
\Box'^2 \Psi' = \gamma \left( \Box^2\Psi + \frac{\beta}{\mu_0 c} \Box^2\Theta_x \right) = \gamma \left( \Box^2\Psi + \frac{\beta}{\mu_0 c} \Box^2\Theta_x \right) = 0.
\]
Similarly, the components of the electric vector potential satisfy:
\begin{align*}
\Box'^2\Theta'_x &= \gamma\left(v\mu_0\Box^2\Psi+\Box^2\Theta_x\right) = 0, \\
\Box'^2\Theta'_y &= \Box^2\Theta_y = 0, \\
\Box'^2\Theta'_z &= \Box^2\Theta_z = 0.
\end{align*}
Therefore, we conclude that:
\[
\Box'^2 \Theta'^\mu = \Box^2 \Theta^\mu = 0,
\]
which demonstrates that the components of the electric vector potential also remain invariant under Lorentz transformations.

\section{Green Function of the D'Alambert operator}
\label{greenFunctionAppendix}

Here we demonstrate the construction of the Green function for the D'Alambert operator in an infinite domain. This procedure is standard but is included here briefly for completeness. The D'Alembert operator is given by

\[
\left(-\nabla^2 + \frac{1}{c^2} \frac{\partial^2}{\partial t^2}\right) G(\boldsymbol{r},t) = \delta(\boldsymbol{r})\delta(t).
\]

\begin{figure}[H]
\centering 
\includegraphics[width=0.6\textwidth]{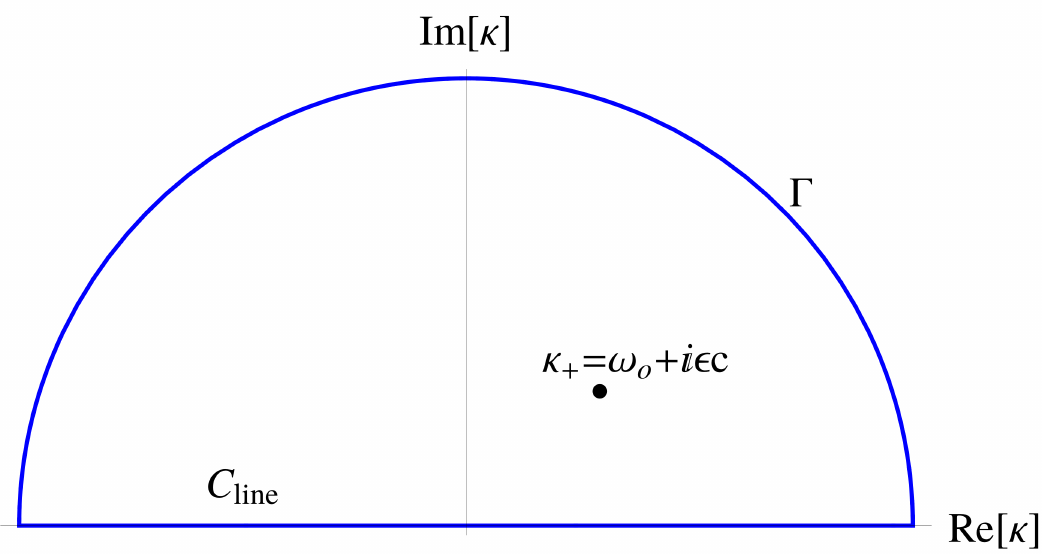}
    \caption[Complex contour of integration.]{Complex contour of integration.  }
\label{theCountourFig}
\end{figure}

We start by taking the Fourier transform:
\[
\hat{G}(\boldsymbol{\kappa},\omega) =\int_{\mathbb{R}^3\cup\mathbb{T}} G(\boldsymbol{r},t) e^{-\boldsymbol{i}(\boldsymbol{\kappa}\cdot\boldsymbol{r}-\omega t)} \frac{d^3\boldsymbol{r}dt}{(2\pi)^4}
\]
where $\boldsymbol{\kappa}$ and $\omega$ are the Fourier space variables. In this Fourier spectral domain, the D'Alembert equation becomes:
\[
-\kappa^2 \hat{G} + \frac{\omega^2}{c^2} \hat{G} = 1 \hspace{0.5cm}\mbox{with}\hspace{0.5cm}\hat{G}(\boldsymbol{\kappa},\omega) = \frac{1}{(\omega/c)^2-\kappa^2}.
\]
Now, we perform the inverse Fourier transform:
\[
G(\boldsymbol{r},t) = \int_{-\infty}^\infty\int_{\mathbb{R}^3} \hat{G}(\boldsymbol{\kappa},\omega) e^{\boldsymbol{i}(\boldsymbol{\kappa}\cdot\boldsymbol{r}-\omega t)} \frac{d^3\boldsymbol{\kappa}d\omega}{(2\pi)^4}=\int_{-\infty}^\infty g(\boldsymbol{r},\omega) e^{-\boldsymbol{i}\omega t} \frac{d\omega}{(2\pi)}\hspace{0.25cm}\mbox{with}\hspace{0.25cm}g(\boldsymbol{r},\omega)=\int_{\mathbb{R}^3} \hat{G}(\boldsymbol{r},t) e^{\boldsymbol{i}\boldsymbol{\kappa}\cdot\boldsymbol{r}} \frac{d^3\boldsymbol{\kappa}}{(2\pi)^3}.
\]
The $g(\boldsymbol{r},\omega)$ function can be evaluated by writing $\boldsymbol{\kappa}$ in spherical coordinates, with $\boldsymbol{\kappa}\cdot\boldsymbol{r}=\kappa r \cos\theta$, $\kappa_x=\kappa\sin\theta\cos\phi$, $\kappa_y=\kappa\sin\theta\sin\phi$, and $\kappa_z=\kappa\cos\theta$. Therefore, in spatial coordinates,
\[
g(\boldsymbol{r},\omega)=\frac{1}{(2\pi^2)}\int_{0}^{2\pi}\int_{0}^{\pi}\int_{0}^\infty \hat{G}(\boldsymbol{\kappa},t) e^{\boldsymbol{i}\boldsymbol{\kappa}\cdot\boldsymbol{r}} \kappa^2\sin\theta d\phi d\kappa = \frac{1}{(2\pi^2)}\int_{\mathbb{R}} \frac{e^{\boldsymbol{i}\kappa r}}{(\omega/c)^2-\kappa^2} \frac{d\kappa}{\boldsymbol{i}r}
\]
by integrating the angular variables. Now, we consider the contour integration in the complex plane. The integral becomes:
\[
I = \oint_C f(\kappa) d \kappa \hspace{0.5cm}\mbox{with}\hspace{0.5cm} f(\kappa)=\frac{1}{(2\pi^2)} \frac{e^{\boldsymbol{i}\kappa r}}{(\omega/c)^2-\kappa^2} \frac{1}{\boldsymbol{i}r}=\frac{1}{(2\pi^2)} \frac{e^{\boldsymbol{i}\kappa r}}{[(\omega/c)-\kappa][(\omega/c)+\kappa]} \frac{1}{\boldsymbol{i}r}.
\]
This integral is taken over a closed contour $C$, which consists of the real line from $-R$ to $R$ (denoted by $c_{\text{line}}$) and the counterclockwise semicircle $\Gamma$ of radius $R$ centered at the origin, as shown in Fig.~\ref{theCountourFig}. Assuming that $\kappa$ is a complex variable, there is only one pole inside the contour $C$. The poles are located at $\kappa_{\pm} = \pm\omega/c = \pm(\omega_o + \boldsymbol{i}\epsilon c)$, where $\epsilon$ is a positive infinitesimal. Applying the Residue Theorem, we find:

\[
I = \oint_C f(\kappa) d \kappa = 2\pi\boldsymbol{i} \mbox{Res}_{f}(\kappa_{+})  = 2\pi\boldsymbol{i} \lim_{\kappa\rightarrow \omega/c} \left(\kappa-\frac{\omega}{c}\right) f(\kappa)=-\frac{1}{(4\pi)}\frac{e^{\boldsymbol{i}\omega r/c}}{r}.
\]

Taking the limit as $R\rightarrow\infty$ with $\epsilon\rightarrow 0$, we find:
\[
\lim_{R\rightarrow\infty} I =  \lim_{R\rightarrow\infty}\int_{C_{line}} f(\kappa) d \kappa+\lim_{R\rightarrow\infty}\int_\Gamma f(\kappa) d \kappa  = -\frac{1}{(4\pi)}\frac{e^{\boldsymbol{i}\omega r/c}}{r},
\]
where the integral vanishes in the semicircular contour $\Gamma$. Therefore, we have:
\[
\lim_{R\rightarrow\infty} I =  \lim_{R\rightarrow\infty}\int_\Gamma f(\kappa) d \kappa = \frac{1}{(2\pi^2)}\int_{\mathbb{R}} \frac{e^{\boldsymbol{i}\kappa r}}{(\omega/c)^2-\kappa^2} \frac{d\kappa}{\boldsymbol{i}r} = -\frac{1}{(4\pi)}\frac{e^{\boldsymbol{i}\omega r/c}}{r}=g_{+}(\boldsymbol{r},\omega),
\]
for $\epsilon>0$. A similar result can be obtained for $\epsilon<0$, then 
\[
g_{\pm}(\boldsymbol{r},\omega) = -\frac{1}{(4\pi)}\frac{e^{\pm\boldsymbol{i}\omega r/c}}{r}.
\]
This result holds for $\epsilon > 0$. A similar result can be obtained for $\epsilon < 0$. Therefore, we have two solutions for the inverse Fourier transform:
\[
G_{\pm}(\boldsymbol{r},t)=\int_{-\infty}^\infty g_{\pm}(\boldsymbol{r},\omega) e^{-\boldsymbol{i}\omega t} \frac{d\omega}{(2\pi)}=-\frac{1}{4\pi r} \int_{-\infty}^\infty e^{\boldsymbol{i}\omega(- t \pm r/c)} \frac{d\omega}{2\pi}.
\]
Now, using the  
\[
\int_{-\infty}^\infty e^{2\pi\boldsymbol{i}\xi t}dt = \delta(\xi)
\]
identity, the following result is obtained
\[
G_{\pm}(\boldsymbol{r},t)=-\frac{1}{4\pi r}\delta\left(-t\pm \frac{r}{c}\right). 
\]

\section{Boundary condition}
\label{BoundaryConditionSectionLabel}
In this section, we will verify that the Green function satisfies the Neumann boundary conditions on the domain boundary $\partial\mathcal{D}$ given by Eq.~(\ref{NeumannCondtionOnGreenFunctionEq}). The domain is defined as $\partial\mathcal{D}=\partial\mathcal{D}_{xy}\cup\partial\mathcal{C}$, where $\partial\mathcal{D}_{xy}=\{(x,y,0) : x\in\mathbb{R}, y\in\mathbb{R}\}$ represents the points of the $xy$-plane. 
Since the Green function and its position derivatives decay as distances from the origin grow, the normal derivatives $\partial_n' G(\boldsymbol{r},\boldsymbol{r}')$ on the surface $\partial\mathcal{C}=\partial\mathcal{D} \setminus  \partial\mathcal{D}_{xy}$, which is located at infinity, tend to zero. We can explicitly show that by considering the following identities:
\[
\frac{\partial}{\partial z'}\frac{1}{|\boldsymbol{r}-\boldsymbol{r}'|} = \frac{(z-z')}{|\boldsymbol{r}-\boldsymbol{r}'|^{3}} \hspace{0.5cm}\mbox{and}\hspace{0.5cm}\left.\frac{\partial}{\partial z'}\frac{1}{|\boldsymbol{r}-\boldsymbol{r}'|}\right|_{z'\longrightarrow -z'} = -\frac{(z+z')}{|\boldsymbol{r}-\boldsymbol{r}'|^{3}} 
\]
therefore
\begin{equation}
\lim_{z' \to 0} \left\{\frac{\partial}{\partial z'}\frac{1}{|\boldsymbol{r}-\boldsymbol{r}'|} +  \left[\frac{\partial}{\partial z'}\frac{1}{|\boldsymbol{r}-\boldsymbol{r}'|}\right]_{z'\longrightarrow -z'}\right\} = 0
\label{auxIdentity1Eq}
\end{equation}
Similarly, we have:
\[
\frac{\partial}{\partial z'} \delta(t'-t_{\boldsymbol{r}}) = \frac{\delta(t'-t_{\boldsymbol{r}})}{t-t_{\boldsymbol{r}}}\frac{\partial t_{\boldsymbol{r}}}{\partial z'} = \frac{\delta(t'-t_{\boldsymbol{r}})}{c(t-t_{\boldsymbol{r}})}\frac{z-z'}{|\boldsymbol{r}-\boldsymbol{r}'|} \hspace{0.5cm}\mbox{and}\hspace{0.5cm}\left.\frac{\partial}{\partial z'} \delta(t'-t_{\boldsymbol{r}})\right|_{z'\longrightarrow -z'} = -\frac{\delta(t'-t_{\boldsymbol{r}})}{c(t-t_{\boldsymbol{r}})}\frac{z+z'}{|\boldsymbol{r}-\boldsymbol{r}'|}
\]
then
\[
\lim_{z' \to 0} \frac{\partial}{\partial z'}\delta(t'-t_{\boldsymbol{r}}) = - \lim_{z' \to 0} \left.\frac{\partial}{\partial z'}\delta(t'-t_{\boldsymbol{r}})\right|_{z'\longrightarrow -z'}.
\]
Additionally, we have:
\[
\lim_{z' \to 0} \frac{1}{|\boldsymbol{r}-\boldsymbol{r}'|} = \lim_{z' \to 0} \left.\frac{1}{|\boldsymbol{r}-\boldsymbol{r}'|}\right|_{z'\longrightarrow -z'}
\]
resulting in:
\begin{equation}
\lim_{z' \to 0} \left\{ \frac{1}{|\boldsymbol{r}-\boldsymbol{r}'|}\frac{\partial}{\partial z'}\delta(t'-t_{\boldsymbol{r}}) + \left[\frac{1}{|\boldsymbol{r}-\boldsymbol{r}'|} \frac{\partial}{\partial z'}\delta(t'-t_{\boldsymbol{r}})\right]_{z'\longrightarrow -z'} \right\} = 0.
    \label{auxIdentity2Eq}
\end{equation}
With these identities in place, we can calculate the normal derivative of the Green function as follows:

\begin{align*}
\lim_{z' \to 0} \frac{\partial}{\partial z'}G_N(\boldsymbol{r},\boldsymbol{r}') &= -\frac{1}{4\pi}\lim_{z' \to 0} \frac{\partial}{\partial z'}\left\{ \frac{\delta(t'-t_{\boldsymbol{r}})}{|\boldsymbol{r}-\boldsymbol{r}'|} + \left[ \frac{\delta(t'-t_{\boldsymbol{r}})}{|\boldsymbol{r}-\boldsymbol{r}'|} \right]_{z'\longrightarrow -z'}\right\} \\
=  -\frac{1}{4\pi}  \lim_{z' \to 0} & \left\{ \frac{1}{|\boldsymbol{r}-\boldsymbol{r}'|}\frac{\partial}{\partial z'} \delta(t'-t_{\boldsymbol{r}}) + \delta(t'-t_{\boldsymbol{r}}) \frac{\partial }{\partial z'} \frac{1}{|\boldsymbol{r}-\boldsymbol{r}'|}  \right.\\
& + \left. \left[\frac{1}{|\boldsymbol{r}-\boldsymbol{r}'|}\frac{\partial}{\partial z'} \delta(t'-t_{\boldsymbol{r}}) + \delta(t'-t_{\boldsymbol{r}}) \frac{\partial }{\partial z'} \frac{1}{|\boldsymbol{r}-\boldsymbol{r}'|}\right]_{z\longrightarrow -z'}  \right\}\\
&=-\frac{1}{4\pi} 
\lim_{z' \to 0} \left\{ \frac{1}{|\boldsymbol{r}-\boldsymbol{r}'|}\frac{\partial}{\partial z'}\delta(t'-t_{\boldsymbol{r}}) + \left[\frac{1}{|\boldsymbol{r}-\boldsymbol{r}'|} \frac{\partial}{\partial z'}\delta(t'-t_{\boldsymbol{r}})\right]_{z'\longrightarrow -z'} \right\} \\
&-\frac{1}{4\pi} \lim_{z' \to 0}\delta(t'-t_{\boldsymbol{r}}) \left\{\frac{\partial}{\partial z'}\frac{1}{|\boldsymbol{r}-\boldsymbol{r}'|} +  \left[\frac{\partial}{\partial z'}\frac{1}{|\boldsymbol{r}-\boldsymbol{r}'|}\right]_{z'\longrightarrow -z'}\right\} \\
&=0
\end{align*}
where we have used Eqs.~(\ref{auxIdentity1Eq}) and (\ref{auxIdentity2Eq}).  

\end{appendices}








\end{document}